\documentclass{aa}  
\newcommand\Tstrut{\rule{0pt}{2.6ex}}         
\newcommand\Bstrut{\rule[-0.9ex]{0pt}{0pt}}  
\usepackage{graphicx}
\usepackage{natbib}
\bibpunct{(}{)}{;}{a}{}{,} 
\usepackage{txfonts}
\usepackage{color}
\usepackage{multirow}
\usepackage[switch]{lineno}
\usepackage{hyperref}
\newcommand{\msun}{\ensuremath{M_{\odot}}}
\newcommand{\lum}{\ensuremath{\mathrm{erg\,s^{-1}}}}
\newcommand{\fermi}{\textit{Fermi}\xspace}
\newcommand{\nustar}{\textit{NuSTAR}\xspace}
\newcommand{\xmm}{\textit{XMM-Newton}\xspace}
\newcommand{\swift}{\textit{Swift}\xspace}
\newcommand{\phflux}{\ensuremath{\mathrm{ph\,cm^{-2}\,s^{-1}}}}
\newcommand{\ergflux}{\ensuremath{\mathrm{erg\,cm^{-2}\,s^{-1}}}}
\newcommand{\gm}{\ensuremath{\gamma}}

\graphicspath{{./plots/}}

\begin{document}

   \title{The first GeV flare of the radio-loud narrow-line Seyfert 1 galaxy PKS\,2004$-$447}
   
      \titlerunning{The first GeV flare of PKS\,2004$-$447}
      \authorrunning{Gokus et al.}

   \author{Andrea Gokus\inst{1,2},
          V.~S.~Paliya\inst{3},
          S.~M.~Wagner\inst{2},
          S.~Buson\inst{2},
          F.~D'Ammando\inst{4},
          P.~G.~Edwards\inst{5},
          M.~Kadler\inst{2},
          M.~Meyer\inst{6},
          R.~Ojha\inst{7,8,9},
          J.~Stevens\inst{10},
          J.~Wilms\inst{1}
          }

   \institute{Remeis Observatory and Erlangen Centre for Astroparticle
     Physics,
     Universit\"at Erlangen-N\"urnberg,
             Sternwartstr.~7, 96049 Bamberg, Germany\\
              \email{andrea.gokus@fau.de}
         \and Lehrstuhl f\"ur Astronomie, Universit\"at W\"urzburg,
              Emil-Fischer-Stra{\ss}e 31, 97074 W\"urzburg, Germany
        \and Aryabhatta Research Institute of Observational Sciences (ARIES), Manora Peak, Nainital 263001, India\\
            \email{vaidehi.s.paliya@gmail.com}
        \and Istituto di Radioastronomia - INAF, Via P. Gobetti 101, 40129 Bologna, Italy
        \and CSIRO Astronomy and Space Science, P.O. Box 76, Epping NSW 1710, Australia
        \and Friedrich-Alexander Universit\"at Erlangen-N\"urnberg, Erlangen Centre for Astroparticle Physics, Erwin-Rommel-Str. 1, 91058 Erlangen, Germany
        \and NASA Goddard Space Flight Center, Greenbelt, MD 20771, USA
        \and Catholic University of America, Washington, DC 20064, USA
        \and University of Maryland, Baltimore County, 1000 Hilltop Cir, Baltimore, MD 21250, USA
        \and CSIRO Astronomy and Space Science, 1828 Yarrie Lake Road, Narrabri NSW 2390, Australia
            }
   \date{Received September 9, 2020; accepted February 20, 2021}
 
  \abstract
   {On 2019 October 25, the \textit{Fermi}-Large Area Telescope observed the first ever $\gamma$-ray flare from the radio-loud narrow-line Seyfert 1 galaxy PKS\,2004$-$447 ($z=0.24$). Prior to this discovery, only four sources of this type had shown a flare at Gigaelectronvolt energies.}
   {We report on follow-up observations in the radio, optical-UV, and
     X-ray bands that were performed by ATCA, the \textit{Neil Gehrels Swift} observatory, \textit{XMM-Newton}, and \textit{NuSTAR}, respectively, and analyse these multi-wavelength data with a one-zone leptonic model in order to understand the physical mechanisms that were responsible for the flare.}
   {We study the source's variability across all energy bands and
     additionally produce $\gamma$-ray light curves with different
     time binnings to study the variability in \gm-rays on
     short timescales during the flare. We examine the combined X-ray
     spectrum from 0.5--50\,keV by describing the spectral shape with
     an absorbed power law. We analyse multi-wavelength datasets
     before, during, and after the flare and compare these with a low
     activity state of the source by modelling the respective spectral
     energy distributions (SEDs) with a one-zone synchrotron inverse Compton
     radiative model. Finally, we compare the variability and the SEDs 
     to \gm-ray flares previously observed from other $\gamma$-loud 
     narrow-line Seyfert 1 galaxies.}
   {At \gm-ray energies (0.1--300\,GeV) the flare reached a maximum flux of $(1.3\pm0.2)\times10^{-6}$~ph~cm$^{-2}$~s$^{-1}$ in daily binning and a total maximum flux of $(2.7\pm0.6)\times10^{-6}$~ph~cm$^{-2}$~s$^{-1}$ when a 3-hour binning was used. With a photon index of $\Gamma_{0.1-300\mathrm{GeV}}=2.42\pm0.09$ during the flare, this corresponds to an isotropic \gm-ray luminosity of $(2.9\pm0.8)\times10^{47}\,\mathrm{erg}\,\mathrm{s}^{-1}$. The \gm-ray, X-ray, and optical-UV light curves that cover the end of September to the middle of November show significant variability, and we find indications for flux-doubling times of $\sim 2.2$~hours at \gm-ray energies. The soft X-ray excess, which is observed for most narrow-line Seyfert 1 galaxies, is not visible in this source. During the flare, the SED exhibits large Compton dominance. While the increase in the optical-UV range can be explained by enhanced synchrotron emission, the elevated \gm-ray flux can be accounted for by an increase in the bulk Lorentz factor of the jet, similar to that observed for other flaring \gm-ray blazars.}


   \maketitle
%
\section{Introduction}
Narrow-line Seyfert 1 (NLSy\,1) objects are typically located in the
centres of spiral galaxies. They differ from normal Seyfert~1
galaxies through their unusual narrow lines originating from 
the broad line region (FWHM (H$\beta)\leq2000\,\mathrm{km\,s}^{-1}$; \citealt{osterbrock}). 
These lines are proportionally strong with regard to the
forbidden [\ion{O}{iii}] $\lambda5007$\AA\ line, with a 
flux ratio of [\ion{O}{iii}]/H$\beta \leq3$. 
Surprisingly, a small percentage ($<7\%$) of NLSy\,1 galaxies are found
to be radio-loud \citep[e.g.][]{komossa2006,2018MNRAS.480.1796S}. When the
\textit{Fermi} Large Area Telescope (LAT) detected $\gamma$-ray
emission from the NLSy\,1 galaxy PMN\,J0948+0022 \citep{first_gNLSy1}
in 2009, these objects became the third class of
active galactic nuclei (AGNs) to be detected at \gm-ray energies. 
This discovery was soon followed by the
detection of three more $\gamma$-NLSy\,1 galaxies based on data accumulated
over one year \citep{first_few_gNLS1}.

The second data release (DR2) of the Fourth \fermi-LAT source 
catalogue \citep[4FGL;][]{fermi_4fgl} reports nine \gm-ray detected 
NLSy1 galaxies, though a few other studies have proposed more 
identifications \citep[see, e.g., ][]{2018rnls.confE..20C, 2018MNRAS.481.5046R, 2018ApJ...853L...2P}.
To distinguish them from non-\gm-ray detected NLSy\,1s, we refer to them
as \gm-NLSy\,1s. Among the small sample of
$\gamma$-NLSy\,1 galaxies, four sources have shown at least one GeV
flare with a luminosity comparable to blazar flares in the first
11 years of \textit{Fermi} observations. These sources are
1H\,0323+342 \citep{paliya2014}, SBS\,0846+513 \citep{dammando2012},
PMN\,J0948+0022 \citep{foschini2011, dammando2015}, and PKS\,1502+036
\citep{paliya2016,dammando_PKS1502}. Gamma-ray variability of days or
weeks is a common feature of blazars
\citep[e.g.][]{abdo_blazarvariability}. Furthermore, the more
powerful blazar class of flat-spectrum radio quasars (\hbox{FSRQs}) appears
to be more variable than the less luminous BL Lacs
\citep{new_variability_blazars}. Flaring \gm-NLSy\,1 galaxies therefore
constitute a very interesting target for observations during states of high
activity since they seem to be scarce and so rarely show flares.

As postulated, for example by \citet{peterson2000}, central black holes
in NLSy\,1 galaxies exhibit high accretion rates for radio-quiet
and radio-loud NLSy 1 galaxies alike; however, there is an ongoing 
debate about the mass of their central engines. While deriving
the mass via virial methods results in fairly low black hole masses of
$M_{\mathrm{BH}}<10^8$\msun~\citep[e.g.][]{grupe_mathur_2004,deo2006,jarvela2017},
other methods suggest that black hole masses of NLSy\,1s are comparable to those in FSRQs \citep[e.g., ][]{decarli2008, marconi2008, viswanath2019}. Detailed discussions on this topic have
recently been summarised by \citet{dammando2019} and
\citet{2019ApJ...872..169P}.

Because Seyfert galaxies generally belong to the class of radio-quiet
AGNs and are not detected in $\gamma$-rays, \gm-NLSy\,1s seem to
contradict the AGN unification scheme \citep{1995PASP..107..803U}.
Their radio and \gm-ray emission suggests that relativistic jets are
present in these systems, indicating possible evolutionary processes
within these sources. Additionally, the nature of their host galaxies remains
an open question.
\citet{hostPKS1502_dammando2018}, for instance, found indications for an elliptical
host galaxy for the \gm-NLSy\,1 PKS\,1502+036, although
\citet{NLSy1hosts_2020} stated that a disk-like host fits their
observation better. A systematic study by the latter authors indicates
that the hosts of radio-loud NLSy\,1s are preferentially disk galaxies,
with a spiral galaxy suggested as the host for PKS\,2004$-$447. 
Furthermore, high-resolution near-infrared imaging of some of the 
NLSy1s has also revealed ongoing galaxy mergers, thus suggesting a 
pivotal role played by such mergers in triggering the jet launching 
\citep[see, e.g.,][]{2020ApJ...892..133P}. Therefore, given the 
small sample of known \gm-ray NLSy 1 galaxies and their unclear nature, 
each new study of one of these objects can help improve our understanding
of the underlying physical properties both in regard to their probable link 
to AGN evolution and to their role within the unified scheme of AGNs.

The most recent $\gamma$-ray flare was detected from PKS\,2004$-$447
\citep[$z=0.24$;][]{atel_13229}. This is the first Gigaelectronvolt flare
observed from this AGN. The source showed a $\gamma$-ray flux of
$(1.1\pm0.2)\times10^{-6}$~ph~cm$^{-2}$~s$^{-1}$ in the
$0.1 - 300$~GeV energy range on 2019 October 25
and continued to stay at a high activity level on the following day.
Its flux during the flare was a factor of $\sim$55 higher than its
average flux as reported in the 4FGL catalogue \citep{atel_13229}.
PKS\,2004$-$447 was one of the first \gm-NLSy\,1 galaxies seen by \textit{Fermi}/LAT 
but has never before shown an outburst comparable to blazar flares. Additionally, 
this AGN is of a somewhat mysterious nature as it does not show typical features compared
to other similarly classified sources. In the X-rays it lacks a soft excess 
\citep[see e.g.][]{gallo2006, 2015MNRAS.453.4037O, kreikenbohm2016}, which is usually
common in the spectra of NLSy\,1 galaxies. In the radio band, PKS\,2004$-$447 shows only 
a little extended emission and a steep spectrum, suggestive of a compact-steep-spectrum (CSS) 
object \citep{2001ApJ...558..578O, gallo2006, schulz2016}. As shown by \citet{schulz2016},
this behaviour is unique among the small \gm-NLSy\,1 sample, and it is also extremely 
rare for \gm-loud AGNs. There are only five CSS sources reported in the Fourth 
\textit{Fermi}/LAT AGN catalogue \citep[4LAC; ][]{4lac}. 

We have carried out a multi-frequency campaign to study this first
\gm-ray flaring event from PKS\,2004$-$447, including observations
from \textit{NuSTAR}, \textit{Swift}, \textit{XMM-Newton}, and the 
Australia Telescope Compact Array (ATCA).
In this paper we present the findings and conclusions of this dataset.
We describe the data reduction
in Section~\ref{sec:datareduction}. We present the results of our
variability analysis in Section~\ref{sec:results-var} and the results of
the X-ray analysis in Section~\ref{sec:results-xray}, and we describe the
model for the spectral energy distribution (SED) in Section~\ref{sec:results-sed}. 
We discuss our findings in
Section~\ref{sec:discussion} and conclude in
Section~\ref{sec:conclusion}. We adopt a flat cosmology of
$H_0= 67.8\,\mathrm{km}\,\mathrm{s}^{-1}\,\mathrm{Mpc}^{-1}$,
$\Omega_{\lambda}=0.692$, and $\Omega_{\mathrm{M}}=0.308$
\citep{planck2016}.

\section{Observations and data reduction}\label{sec:datareduction}

   \begin{figure*}
   \centering
        \includegraphics[width=17cm]{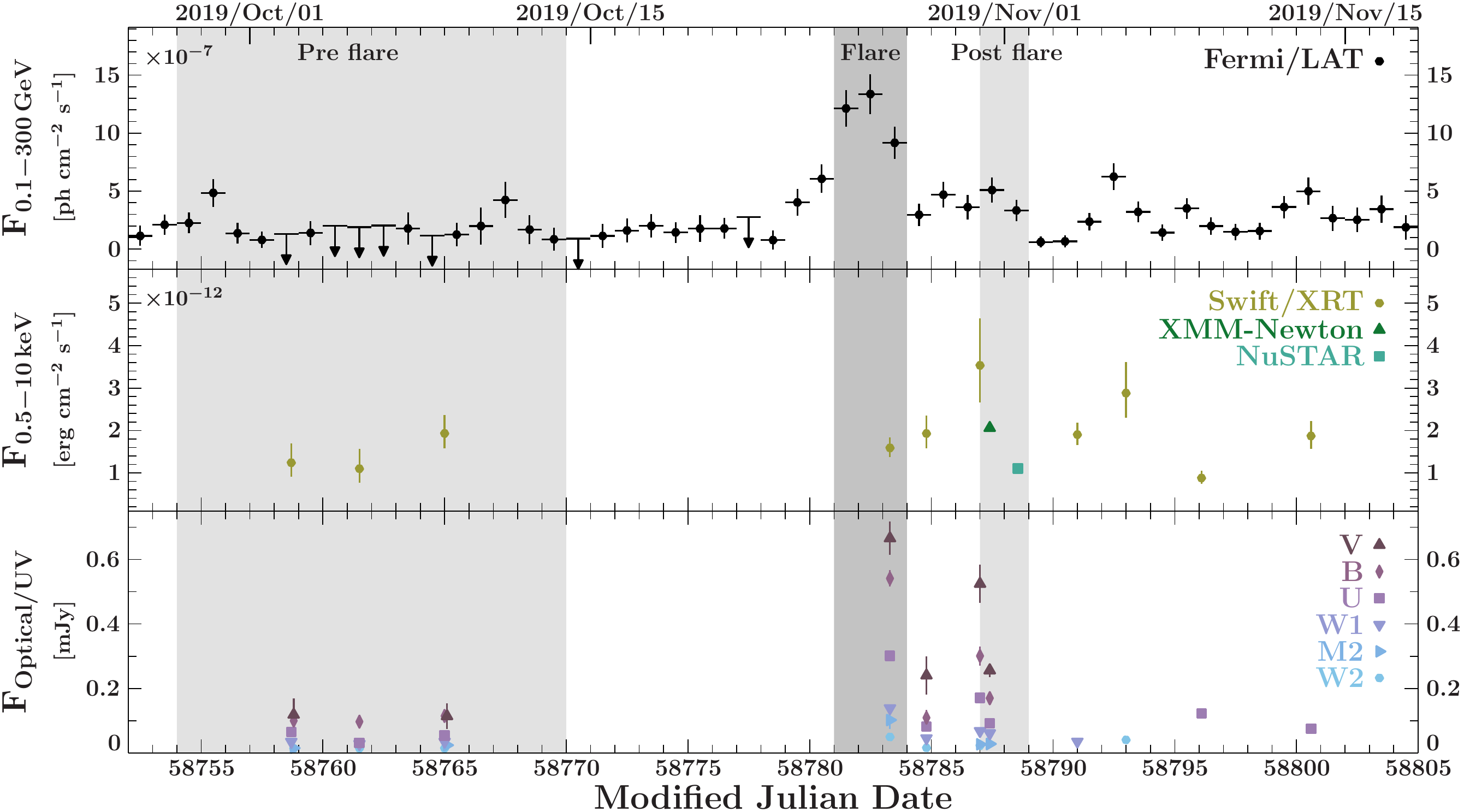}
        \caption{Light curves in the $\gamma$-ray (top), X-ray (middle), and UV/optical (bottom)
                regimes from 2019 September 27 until 2019 November 19.The \textit{Fermi}-LAT light curve shows the daily binned flux of the bins with $\mathrm{TS}\geq1$. LAT light curve bins with $\mathrm{TS}<1$ are represented as $2\sigma$ upper limit arrows. All errors represent the $1\sigma$ uncertainties. For \textit{NuSTAR}, the flux is interpolated down to 0.5~keV. The dark-grey shadowed region marks the time range that is used for the flare SED, while the pre- and post-flare time ranges are shown in light grey.}
        \label{fig:lightcurves}
    \end{figure*}

\subsection{Fermi Large Area Telescope observations}
The LAT on board the \textit{Fermi} satellite
is a pair-conversion telescope. It has been in operation since 2008
\citep{fermi_instr}. LAT's all-sky monitoring strategy provides a full
coverage of the $\gamma$-ray sky every 3.2\,h, that is two
\textit{Fermi} orbits. Our reduction of the \textit{Fermi}-LAT data
follows the standard data reduction
process\footnote{\href{https://fermi.gsfc.nasa.gov/ssc/data/analysis/documentation/}{https://fermi.gsfc.nasa.gov/ssc/data/analysis/documentation/}}
and uses the \texttt{Science Tools v11r04p00}. We extract those events
suitable for an analysis\footnote{We use SOURCE class events and set the following flags: 
\texttt{(DATA\_QUAL>0)\&\&(LAT\_CONFIG==1)}} with energies in the range from 100\,MeV to
300\,GeV in a region of interest (ROI) of $15^\circ$ centred at the
4FGL position of PKS\,2004$-$447. We ignore all events with zenith
angles $\ge90^\circ$, in order to exclude $\gamma$-rays originating from
Earth-limb effects. We use the post-launch instrument response
function \texttt{P8R3\_SOURCE\_V2}, \texttt{gll\_iem\_v07} as the
Galactic diffuse model and \texttt{iso\_P8R3\_SOURCE\_V2\_v1} to model the 
isotropic diffusion emission\footnote{The background models are available
at \href{https://fermi.gsfc.nasa.gov/ssc/data/access/lat/BackgroundModels.html}{https://fermi.gsfc.nasa.gov/ssc/data/access/lat/BackgroundModels.html}}.
We use a maximum likelihood analysis to optimise our model parameters and determine
the significance of the modelled $\gamma$-ray signal
via the test statistic $\mathrm{TS}=2\Delta\log(\mathcal{L})$, 
where $\mathcal{L}$ is the likelihood function that represents 
the difference between models with and without a point source at the source coordinates
\citep{mattox}. Our model includes all 4FGL sources within $20^\circ$
of PKS\,2004$-$447. Following the 4FGL, we model the spectrum of 
PKS\,2004$-$447 with a logarithmic parabola and 
discuss the significance of the spectral curvature during the time of the flare
in Section \ref{sec:results-var}. For all sources within $3^\circ$ that have
$\mathrm{TS}\geq10$, as well as for the isotropic and Galactic diffuse
components, we leave the normalisation free to vary but keep the spectral
parameters as reported in the 4FGL. For the sources not fulfilling
these requirements, the normalisation is set to their respective 4FGL value as well.

Data used for compiling the low-state SED are centred in time on \textit{Swift} and \textit{XMM-Newton}
observations performed in 2012 March and May, respectively
\citep[see][]{kreikenbohm2016}. They cover 24 months from
2011 May 1 through 2013 May 1.

We compute a daily-binned light curve from 2019 September 26 to 2019
November 19, shown in Fig.~\ref{fig:lightcurves}, and keep
the parameters of all sources in the ROI fixed to the values derived 
by the analysis over this time range.
For a deeper investigation of the $\gamma$-ray
variability in the source around the time of the \gm-ray flare,
we go to smaller time binnings (12\,h, 6\,h, and 3\,h). 
We generate these light curves similarly to the daily binned light curve, but 
over a slightly shorter time range, from 2019 October 10 to 2019 November 14.
Uncertainties for all \textit{Fermi}-LAT light curves are shown 
at the $1\sigma$ level.

\subsection{Neil Gehrels Swift Observatory observations}
Following our detection of the flare of PKS\,2004$-$447 on 2019
October 25 \citep{atel_13229}, we triggered a target of opportunity
observation with the \textit{Swift} satellite \citep{swift}, which was
performed on 2019 October 27 \citep{atel_13233}. This observation was
followed by several follow-up observations on 2019 October 28 and 30 and
2019 November 4, 6, 9, and 13. The \textit{Swift}
observations were performed in photon-counting mode. In order to clean
the data and create calibrated event files we used the standard
filtering methods and \texttt{xrtpipeline}, as distributed in the
HEASOFT (v6.26) package. The spectrum of the source was accumulated
from a circular region with a radius of $35''$. The background region
was defined by an annulus with an inner radius of $50''$ and an outer
radius of $150''$ at the same coordinates as the source region.

To derive the source fluxes and describe the spectral shape, we use
the Interactive Spectral Interpretation System \citep[ISIS, Version
1.6.2-40][]{isis}. Throughout this paper, we describe the absorption
in the interstellar medium using \texttt{vern} cross sections
\citep{vern} and \texttt{wilm} abundances \citep{wilms_tbabs}. We use
C-statistics \citep{cash} and estimate all uncertainties at 68\%
confidence (1 $\sigma$).
The source spectra are binned after the algorithm described by \citet{optimalbinning}
in order to ensure optimal binning. 
We adopt an absorbed power law (\texttt{tbabs*powerlaw}) to model each spectrum.
The Galactic \ion{H}{i} column density,
$N_\mathrm{H,Gal}=2.97\times10^{20}\,\mathrm{cm}^{-2}$, is taken from
the \ion{H}{i} 4$\pi$ survey \citep[HI4PI; ][]{HI4PI}, modelled with
\texttt{tbabs} \citep{wilms_tbabs}, and kept fixed during the fit.
The observations confirm a high state of the X-ray flux compared to
previous X-ray observations \citep[an overview of all X-ray
observations between 2004 and 2012 is given by][]{kreikenbohm2016}.
The results are listed in Table~\ref{tab:xrayobs} together with the results from the \textit{XMM-Newton} and \textit{NuSTAR} data analysed in this work. 
For the X-ray light curve we analyse each
\textit{Swift}/XRT observation individually. For building SEDs we
stack all \textit{Swift} observations that fall into the time interval
considered. 

\begin{table*}
    \centering
    \caption{Results from the analysis of the individual X-ray observations by \textit{Swift}/XRT (S), \textit{XMM-Newton}/MOS+pn (X) and \textit{NuSTAR}/FPMA+B (N). We report unabsorbed fluxes in units of 10$^{-12}$ \ergflux. NB: The photon index reported for \textit{NuSTAR} is the index for the full \textit{NuSTAR} energy range from 3 to 79\,keV and the flux is extrapolated down to 0.5\,keV.}
    \begin{tabular}{ccccccc}
        \hline\hline
         ObsDate & Instrument & ObsID & Net Exposure [ks] & $\Gamma_{0.5-10~\mathrm{keV}}$ & Flux$_{0.5-10~\mathrm{keV}}$ & Statistics (C-stat./dof) \Tstrut\Bstrut\\
         \hline
         2019-10-02 & S & 00081881003 & 1.2 & $1.3\pm0.4$ & $1.2^{+0.5}_{-0.3}$ & 55.72/45\Tstrut\\
         2019-10-05 & S & 00081881004 & 0.9 & $1.4\pm0.4$ & $1.1^{+0.5}_{-0.3}$ & 27.30/45\Tstrut\\
         2019-10-09 & S & 00081881005 & 2.0 & $1.02\pm0.21$ & $1.9\pm0.4$ & 40.03/45\Tstrut\\
         2019-10-27 & S & 00032492020 & 2.9 & $1.62\pm0.18$ & $1.59^{+0.25}_{-0.22}$ & 59.59/46\Tstrut\\
         2019-10-28 & S & 00032492021 & 2.0 & $1.14\pm0.21$ & $2.0^{+0.4}_{-0.3}$ &  46.34/45\Tstrut\\
         2019-10-30 & S & 00032492022 & 1.6 & $0.69^{+0.29}_{-0.30}$ & $3.5^{+1.1}_{-0.9}$ & 61.51/45\Tstrut\\
         2019-10-31 & X & 0853980701 & 11.2 & $1.424\pm0.024$ & $2.06\pm0.05$ & 97.23/80\Tstrut\\
         2019-11-01 & N & 90501649002 & 30.1 & $1.31\pm0.05$ & $1.10\pm0.04$ & 357.54/331\Tstrut\\
         2019-11-04 & S & 00032492024 & 3.6 & $1.31\pm0.15$ & $1.90^{+0.29}_{-0.25}$ & 61.64/46\Tstrut\\
         2019-11-06 & S & 00032492025 & 0.7 & $1.79^{+0.26}_{-0.25}$ & $2.8^{+0.7}_{-0.6}$ &  33.31/45\Tstrut\\
         2019-11-09 & S & 00032492026 & 3.4 & $1.97\pm0.23$ & $0.88^{+0.17}_{-0.14}$ & 38.50/45\Tstrut\\
         2019-11-13 & S & 00032492027 & 2.5 & $1.27\pm0.19$ & $1.9^{+0.4}_{-0.3}$ & 50.87/45\Tstrut\Bstrut\\
         \hline
    \end{tabular}
    
    \label{tab:xrayobs}
\end{table*}

Simultaneously to the XRT, the Ultraviolet/Optical Telescope (UVOT) on
board \textit{Swift} was also observing the source. We use this
instrument to derive optical and ultraviolet fluxes. The data are
reduced using the standard procedures with a source region of $5''$
and a background annulus with an inner radius of $7''$ and an outer
radius of $21''$. The optical-UV fluxes shown in this paper are
dereddened via the $E(\mathrm{B}-\mathrm{V})$ correction using the Fitzpatrick parametrisation \citep{fitzpatrick}. The magnitude values are converted to flux units using the unfolding procedure implemented in \texttt{ISIS}, which is a model-independent approach described by \citet{isis_unfolding}.
The optical-UV light curve is shown in Fig.~\ref{fig:lightcurves}.

\subsection{\textit{XMM-Newton} observations}
In addition to the \textit{Swift} monitoring, we performed an
\textit{XMM-Newton} ToO observation on 2019 October 31 with an
exposure time of 11~ks (ObsID: 0853980701). Archival data taken during
the low state of the source were obtained from an observation in 2012
May, which has been discussed in detail by \citet{kreikenbohm2016}.
The observations by \textit{XMM-Newton} \citep{jansen2001} were performed 
with both the PN \citep{epic-pn} and the MOS \citep{epic-mos} CCD arrays 
of the European Photon Imaging Camera (EPIC). The observation of optical-UV 
emission was conducted with the Optical Monitor \citep[OM; ][]{xmm-om}. 

The observation with the EPIC was performed in the Small Window Mode
with a thin filter. We use standard methods of the \textit{XMM-Newton}
Science Analysis System (SAS, Version 18.0) to process the observation
data files, and to create calibrated event lists and images. We
extract the source spectrum and a light curve for an energy range
from 0.5\,keV to 10\,keV from a circular region of $35''$ radius
around the source. The background is taken from a circle with a
radius of $60''$. For both the source and the background spectra we
extract the single and double event patterns for the EPIC-pn detector
and all events for the EPIC-MOS detectors. Pile-up is negligible in
the observation. We fit the spectra of the EPIC-MOS and EPIC-pn detectors
simultaneously with an absorbed power law, while using the optimal binning
approach. The result is listed in
Table~\ref{tab:xrayobs} together with the results from the analysis of
\textit{Swift}/XRT and \textit{NuSTAR} observations. 
The X-ray flux of PKS\,2004$-$447 seen by \textit{XMM-Newton} shortly
after the flare is also part of the X-ray light curve in Fig.~\ref{fig:lightcurves}.

The OM observed the source in the $v$, $b$, $u$, $w1$, and $m2$
filters in imaging mode with an exposure time of 1200\,s, 1200\,s,
1200\,s, 1780\,s, and 2200\,s, respectively. The data were processed
using the SAS task \texttt{omichain} and \texttt{omsource}. For the
count rate to flux conversion we used the conversion factors given in
the SAS watchout dedicated page
\footnote{https://www.cosmos.esa.int/web/xmm-newton/sas-watchout-uvflux.}.
The optical/UV fluxes were dereddened via $E(\mathrm{B}-\mathrm{V}$)
correction, using the same approach as for \textit{Swift}, and are
included in the light curve shown in Fig.~\ref{fig:lightcurves}.

\subsection{NuSTAR observation}
We performed a ToO observation with the \textit{Nuclear Spectroscopic
Telescope Array} \citep[\textit{NuSTAR};][]{nustar} with an exposure of
30\,ks on 2019 November 1 (ObsID:~90501649002). We use standard
methods of the software package \texttt{NUSTARDAS} (Version v1.8.0)
distributed in \texttt{HEASOFT} and the calibration database (CALDB)
20190812 to reduce and extract the data for both Focal Plane Modules~A
and~B (FPMA, FPMB). We use \texttt{nuproducts} to create spectra and
response files. We choose a circular region with $50''$ radius for the
source region, and a circle with $120''$ radius in a source-free
region as the background region.
We use the same binning method as we used for the \textit{Swift}/XRT and
\textit{XMM-Newton} spectra and fit the spectra from FPMA and FPMB 
simultaneously with an absorbed power law from 3 to 79\,keV. The result is
given in Table \ref{tab:sed_par}.
In order to compare the flux directly with the other X-ray observations
in the light curve in Fig.~\ref{fig:lightcurves}, we extrapolate the flux 
down to 0.5\,keV and list this value for the flux in Table \ref{tab:xrayobs}.
Initial modelling of the data shows a slight indication for a spectral
hardening at higher energies that is, however, also compatible with residuals 
caused by slight variations of the background level at the $\sim$10\% level. 
In our final fits we therefore vary the normalisation of the 
background by introducing a multiplicative constant that accounts for
this variation.

\subsection{ATCA observations}
As part of the TANAMI blazar monitoring programme, ATCA has been observing
PKS\,2004$-$447 at multiple radio frequencies since 2009\citep{stevens2012}. The
ATCA is an array of six 22-m diameter radio antennas located in
northern New South Wales, at a latitude of $-30^\circ$ and altitude
237\,m above sea level. Its baselines can be adjusted and its
configuration is typically changed every few weeks. The longest
possible baseline is 6\,km. ATCA receivers can be quickly switched
enabling observations to be made over a large range of frequencies in
a short period of time. For our study, monitoring data between
5.5\,GHz and 40\,GHz are collected for the pre-flare and the flaring
states\footnote{Supplementary data from the C\,007 ATCA
  calibrator programme were used.}. The data consist of snapshot observations of
PKS\,2004$-$447 covering a duration of several minutes, which were calibrated against
the ATCA primary flux calibrator PKS\,1934$-$638. Data reduction is
carried out in the standard manner with the MIRIAD software
package\footnote{http://www.atnf.csiro.au/computing/software/miriad/}.

\section{Results}\label{sec:results}
\begin{figure*}
   \centering
    \includegraphics[width=17cm]{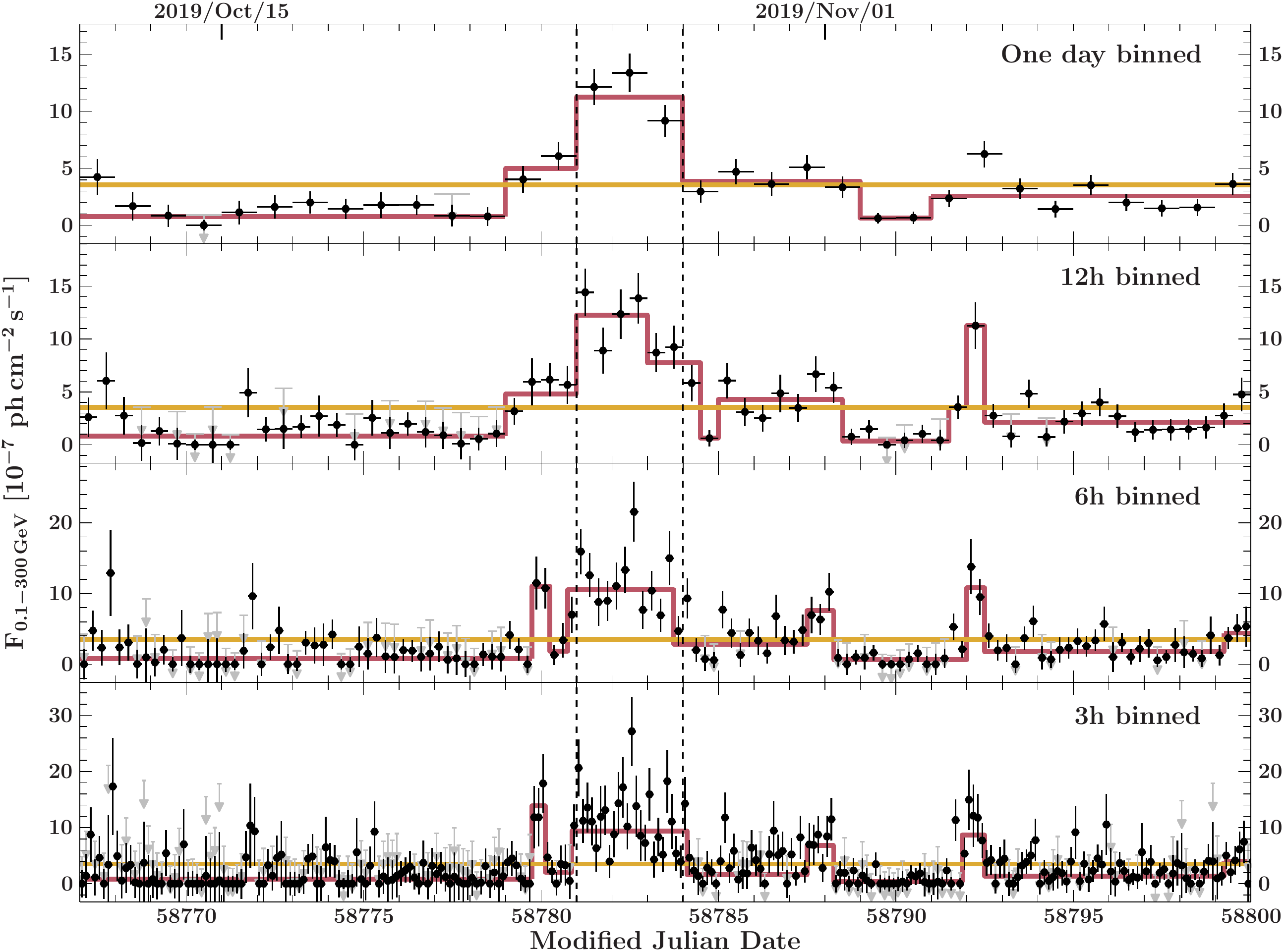}
    \caption{Gamma-ray light curves of PKS\,2004$-$447 during the 2019 GeV flare for the different time binnings of the LAT data. Time bins with $\mathrm{TS}<1$ have an estimated error that was derived using their $1\sigma$ upper limit value. For those bins we show the $2\sigma$ upper limits in grey to visually show how many bins would have been rejected from the analysis otherwise. The Bayesian blocks are shown in red, while the baseline (average flux during the time range from 2019 October 10 to 2019 November 14) is shown in yellow. The dashed lines mark the time range chosen to construct the broadband SED of the flaring state.}
     \label{fig:diffbin}
\end{figure*}
\begin{figure*}
    \centering
    \includegraphics[width=17cm]{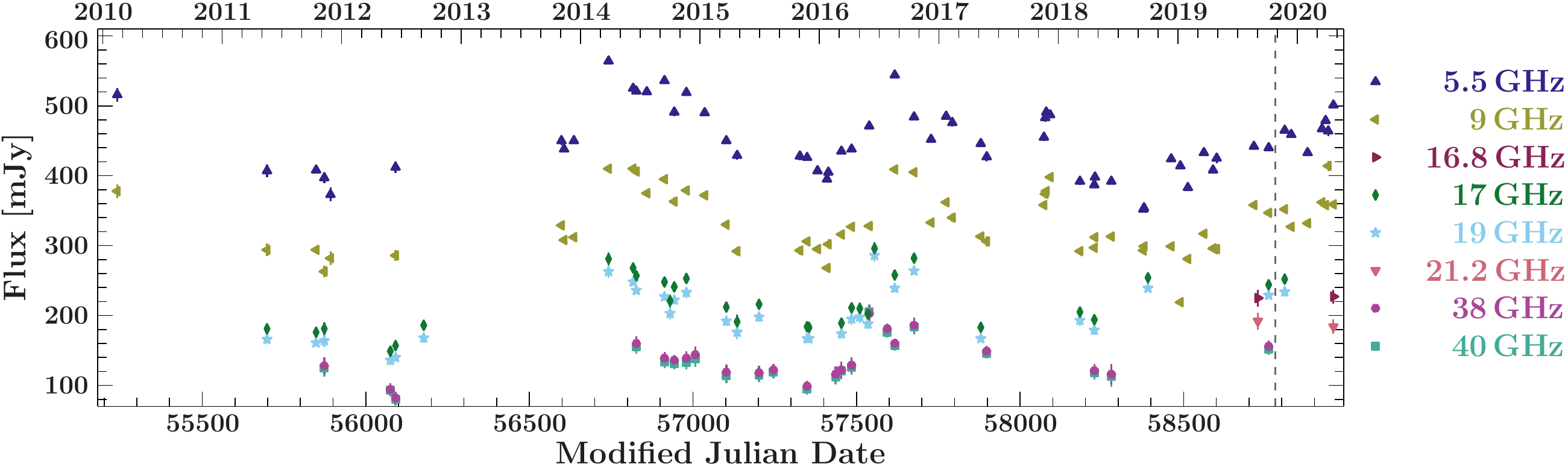}
    \caption{ATCA light curves taken from 2010$-$2020
      in the $\lambda$4-cm (5.5\,GHz, 9\,GHz), $\lambda$15-mm (16.8\,GHz, 17\,GHz,
      19\,GHz, and 21.2\,GHz) and $\lambda$7-mm band (38\,GHz, 40\,GHz).
      The time of the flare is marked by a dashed grey line.}
    \label{fig:atca_lc}
\end{figure*}
\begin{figure}
   \centering
    \includegraphics[width=.85\linewidth]{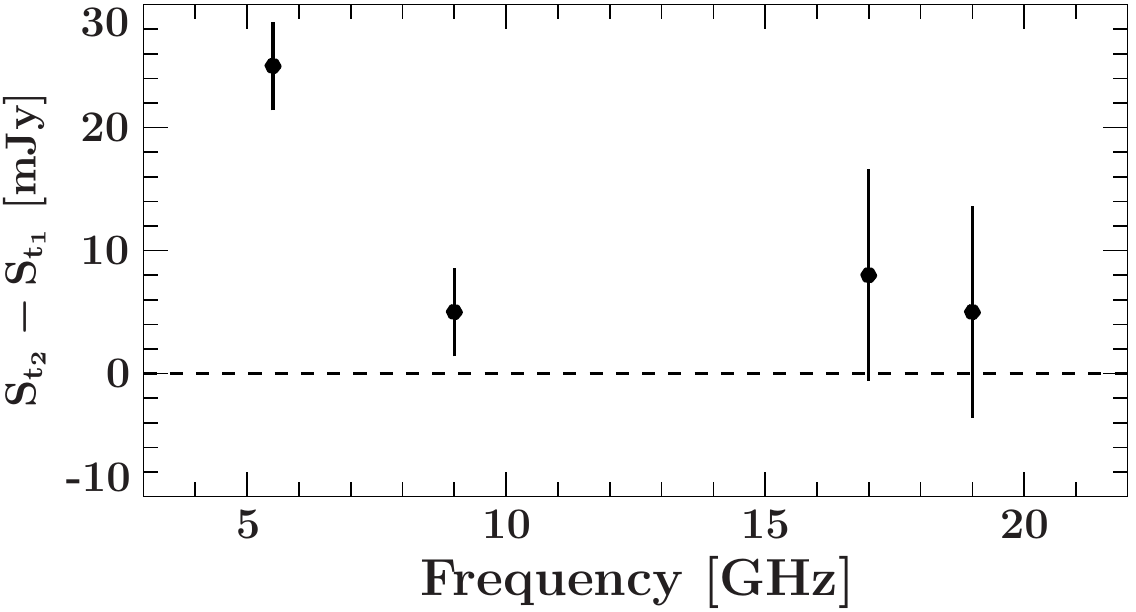}
     \caption{Difference spectrum derived from the radio observations 21\,days before ($t_1$) and 28\,days after ($t_2$) the flare. The dashed line marks the zero change of the flux.}
    \label{fig:radio_diffspec}
\end{figure}

\subsection{Variability}\label{sec:results-var}
Figure~\ref{fig:lightcurves} shows the light curves for PKS\,2004$-$447
based on the daily-binned $\gamma$-ray emission, and individual
X-ray observations by \textit{Swift}/XRT and \textit{XMM-Newton}/EPIC,
and optical-UV observations by \textit{Swift}/UVOT and
\textit{XMM-Newton}/OM. The $\gamma$-ray flux started to rise on 2019
October 23 (MJD~58779). It reached a daily-averaged maximum of
$(1.2\pm0.2)\times10^{-6}$~ph~cm$^{-2}$~s$^{-1}$, which was maintained
over about two days. After that, the flux decreased within two days,
returning to the same flux level as before the flare. In the 3\,h
binned \gm-ray light curves, on 2019 October 26 (MJD~58782.6), we find
a maximum flux of $(2.7\pm0.6)\times10^{-6}$~ph~cm$^{-2}$~s$^{-1}$.
This is the highest \gm-ray flux ever measured for PKS\,2004$-$447.
Together with the spectral index of
$\Gamma_{0.1-300\mathrm{GeV}}=2.42\pm0.09$ measured during the flare,
we derive an isotropic \gm-ray luminosity of
$(2.9\pm0.8)\times10^{47}$~erg~s$^{-1}$. The light
curves binned on different timescales are shown in Fig.~\ref{fig:diffbin}.
For all analyses that follow, we do not include any time bins with $\mathrm{TS} < 4$,
but, for visual purposes, we plot these data points as upper limits in Fig.~\ref{fig:lightcurves} and
Fig.~\ref{fig:diffbin}. 

At X-ray energies (0.5--10\,keV), the flux was highest on 2019
October 30 (MJD~58786), with a flux of
$3.3^{+1.1}_{-0.9}\times10^{-12}$~erg\,cm$^{-2}$\,s$^{-1}$ and a power
law index of $0.8\pm0.3$. The short exposure time of this
observation results in poor constraints on the spectral parameters.
The optical emission in the V, B, and U bands shows strong variations.
The maximum flux occured on 2019 October 27 (MJD~58783), which
coincides with the time of the \gm-ray flare.

To quantify the variability, we first apply a $\chi^2$ test against the
null hypothesis that the emission from PKS\,2004$-$447 is constant in
each energy band.
In the \gm-ray band, we find a null-hypothesis probability of
$p<0.006$ for each of the light curves, regardless of their time binning, thus confirming
variability. With a $p$-value $<0.00001$, the X-ray light curve shown in
Fig.~\ref{fig:lightcurves} exhibits significant variability as well. 
On shorter timescales, however, no significant variability is detected in
either the \textit{XMM-Newton} ($p=0.06$) or the \textit{NuSTAR}
($p\sim1$) data. In the optical-UV band, strong variability
($p < 0.03$) at a level of up to a factor of five compared to 
the flux before the flare is observed with the
maximum roughly coinciding with the \gm-ray flare.

Variability is also seen in the ATCA radio light curves (see
Fig.~\ref{fig:atca_lc}). This is in agreement with earlier work
by \citet{schulz2016}, who discussed the radio variability of
PKS\,2004$-$447 based on TANAMI/ATCA observations between 2010 and
2014 and found moderate variability. Given that only two observations
are located in the time range in which we analysed the \gm- and X-ray
variability, we do not conduct the chi-squared test on these.
Following the ATCA calibrator database
documentation\footnote{\url{https://www.narrabri.atnf.csiro.au/calibrators/calibrator_database_documentation.html}},
we have flagged several epochs that were plotted in
\citet{schulz2016}. We show an updated version of the PKS\,2004$-$447
radio light curve, including data up to early 2020. These data are
presented in Table~\ref{tab:atca_data} of Appendix~A1. The
uncertainties reported are statistical only and do not include any
systematic errors, which in general are known to be smaller than 5\%
in the centimetre bands \citep{tingay2003}.

The radio emission of PKS\,2004$-$447 through 2018 until early 2020
can be characterised by an overall rising trend in all radio bands. 
In the months prior to the \gm-ray flare (marked by the dashed grey
line in Fig.~\ref{fig:atca_lc}), PKS\,2004$-$447 showed a relatively constant flux-density level
of about 440\,mJy at 5.5\,GHz and 350\,mJy at 9\,GHz. Full broadband
radio spectra of PKS\,2004$-$447 were taken on 2019 October 4 and 2019
November 22, namely about 21\,days before and 28\,days after the 2019
October 25 $\gamma$-ray flare. Figure~\ref{fig:radio_diffspec} shows
a difference spectrum, which illustrates the difference between the spectra
derived during each of these two epochs.
While the higher frequencies show only a mild
increase in radio emission after the flare, the 5.5\,GHz emission
increased by about 25\,mJy (${\sim}6$\%). It is not possible to determine
whether this increase is related to the \gm-ray flare. For other
AGNs, delays of a few months have been reported between \gm-ray flares
and subsequent radio flux density increases \citep[e.g.
][]{fuhrmann2014, ramakrishnan2015}.

To look further into the flare behaviour in \gm-rays, the
Bayesian-block algorithm is applied \citep{scargle_BB}.
According to \citet{meyer2019}, a flare can be described as a
group of blocks, which is determined by applying the HOP\footnote{The name
HOP is not an acronym, but taken from the verb 'to hop' to each data 
element’s highest neighbour \citep{hop_algorithm_1998}.} algorithm.
In this algorithm, each Bayesian block that surpasses a certain
baseline is assigned to belong to its highest adjacent block. For this
work, we chose the mean flux of each light curve to represent the baseline flux,
as illustrated in yellow in Fig.~\ref{fig:diffbin}. The total
duration of the flare can then be defined as the time range between
the beginning of the first block and the end of the last block above
the baseline, while the peak is assumed to be located at the centre of
the maximum block. This time range is defined as a
HOP group\footnote{\citet{meyer2019} added an additional criterion
requiring that the maximum block is at least five times above the
average flux in order to single out only the brightest flares, which
we drop in our analysis.}, for which we measure the rise time
$\tau_{\mathrm{rise}}$ from the beginning of the HOP-group to the
peak, and the decay time $\tau_{\mathrm{decay}}$ from the peak to the
end of the HOP-group. We conservatively estimate the error on the 
edge of each Bayesian block to be as big as the binning of each
respective light curve (e.g. $\pm 1$d in daily binning). 
To apply this method to the \textit{Fermi} light
curves of PKS\,2004$-$447, we calculate the Bayesian blocks as 
described by \citet{scargle_BB}, and set the parameter \texttt{ncp\_prior}$=2$. 

A source is not necessarily detected significantly in each light curve bin, hence upper limits on the flux are usually reported (see e.g. the \textit{Fermi}-LAT light curve in Fig.~\ref{fig:lightcurves}) in order to give an indication about the trend of the flux of a source. 
In a standard LAT light-curve analysis, it is not straightforward to deal with data bins that have a low test statistic. Moreover, the number of such low-significance
flux bins typically increases for a finer binning. Specifically, this is problematic for the Bayesian-block point algorithm which assumes that the flux in each bin follows Gaussian statistics. For a low source significance this assumption is not valid. A common approach is to ignore the upper limits altogether as upper limits cannot be inserted as such in the Bayesian-block algorithm, and therefore waive the information contained in data points with low significance, thus biasing the analysis results. 
To avoid this, we take all data into account and calculate best-possible flux values also in the case of low-significance data bins following the standard analysis procedure.
For light-curve bins that have a low significance, a problem that occurs in the determination of the fluxes and their corresponding uncertainties with the Likelihood calculation is that the Likelihood fit does not converge and this can then yield unreasonably small values for the flux uncertainties. This can have a strong influence on the Bayesian-block algorithm. Hence, to avoid this issue, rather than relying on the Likelihood to provide the uncertainties on the flux values, we calculate the 1-sigma upper limits for the flux in the low-flux bins and use the difference between these upper-limit values and the flux returned by the Likelihood as a conservative proxy for the magnitude of the flux uncertainties. In this way our light curve does not exhibit gaps and the Bayesian-block algorithm can be applied to a continuous dataset.

The results from the Bayesian-block algorithm are shown in red in Fig.~\ref{fig:diffbin}. 
Following \citet{meyer2019} we define the flare asymmetry via

\begin{equation}\label{eq:asymmetry}
\centering
    A = \frac{\tau_{\mathrm{rise}}-\tau_{\mathrm{decay}}}{\tau_{\mathrm{rise}}+\tau_{\mathrm{decay}}}.
\end{equation}
Uncertainties are obtained using Gaussian error propagation. The
results are shown in Table~\ref{tab:hop_results}.

\begin{table}
  \centering
  \caption{Flare lengths in days for the different binnings of the \textit{Fermi} light curves. The times $\tau_{\mathrm{rise}}$ and $\tau_{\mathrm{decay}}$ are derived via the HOP algorithm applied on the Bayesian block analysis. $A$ is the asymmetry as defined in Eq. \ref{eq:asymmetry}.}
  \begin{tabular}{lccc}
  \hline\hline
    & $\tau_{\mathrm{rise}}$ [d] & $\tau_{\mathrm{decay}}$ [d] & $A$ \\
    \hline
    Daily & $3.5\pm1$ & $6.5\pm1$ & $-0.30\pm0.15$ \\
    12\,h & $3.0\pm0.5$ & $2.5\pm0.5$ & $0.09\pm0.13$ \\
    6\,h & $1.5\pm0.25$ & $1.5\pm0.25$ & $0.00\pm0.12$ \\
    3\,h & $1.625\pm0.125$ & $1.625\pm0.125$ & $0.00\pm0.05$ \\
    \hline
  \end{tabular}
  \label{tab:hop_results}
\end{table}
  
The asymmetry values depend on the binning size chosen for the light curve:
For the daily-binned light curve, the procedure yields an asymmetry value $A<0$, 
indicating a faster rise than decay of the flare. 
The 12-hour binned light curve resolves more structure and a local dip at MJD 58785 followed by an increased flux level separated from the flare. Due to this the resulting Bayesian blocks indicate a slightly faster decay than rise ($A>0$). The 6- and 3-hour binning, in turn, resolve this to consist of a very short and a longer symmetric flare. We focus on the latter, which lies within the time range chosen to construct the SED of the flaring state as indicated with dashed lines in Fig.~\ref{fig:diffbin}. The properties of this flare and the corresponding higher binnings are reported in Table \ref{tab:hop_results}, but it is important to note that the 6- and 3-hour flares only represent a fraction of the daily and 12-hour one. Furthermore, the 6- and 3-hour binned flares consist of one block only which, by definition, results in $A=0$. Thus, the perceived flare symmetry is most likely due to the analysis procedure and limited sensitivity rather than actual symmetry of the flux behaviour and the true flare shape remains unknown.
Interestingly, nine days after the main flare a second, shorter flare
is identified by the Bayesian-block algorithm in all light curves but the daily-binned one. 
This demonstrates that the $\gamma$-ray variability of the source takes place on sub-day scales. What appears to be one flare in daily binning is shown to consist of three independent flares in 6- and 3-hour binning. Unfortunately, the sensitivity of the instrument is not high enough to fully resolve this structure. In general, care has to be taken in the interpretation of Bayesian flare-duration studies by considering and testing different bin sizes.

To quantify this sub-dayscale variability, we scan all \textit{Fermi} light
curves for significant jumps in flux between adjacent data points and
calculate the minimum doubling and halving times. The most significant
flux difference (${\sim}2.88\sigma$) between adjacent data points is
found in the 6-hour binned light curve at MJD\,58792.0, during the second,
shorter flare. We compute a flux-doubling timescale of
$\tau_{d} = 2.2\pm0.8$ hours, assuming an exponential rise
\citep{zhang1999}.

We search for the presence of spectral curvature in the \gm-ray
spectrum of the brightest state during the flare (MJD 58781-58784) and
obtain the curvature via

\begin{equation}
    \mathrm{TS}_\mathrm{curve} = 2(\mathrm{log}\mathcal{L}(\mathrm{logparabola})-\mathrm{log}\mathcal{L}(\mathrm{powerlaw}))
\end{equation}
from \citet{2FGL}.
Our analysis yields $\mathrm{TS}_\mathrm{curve}=11.66$, providing
tentative evidence for the presence of curvature in the \gm-ray
spectrum.
Although the photon index of $2.42\pm0.09$ measured during the
flare is marginally harder than the average photon index of $2.60\pm0.05$
reported in 4FGL \citep{fermi_4fgl}, the difference is not large enough
to claim that spectral hardening has taken place during the flare.
PKS\,2004$-$447 is significantly detected up to an energy of 3\,GeV during the flare. The slight curvature of the spectrum and the increasing flux threshold for detection are responsible for the non-detection at higher energies. Attenuation of the \gm-ray emission seen by \textit{Fermi}-LAT due to pair production with the extragalactic background light (EBL) is negligible at these energies for the redshift ($z=0.24$) of PKS\,2004$-$447.

\subsection{Analysis of the X-ray spectra}\label{sec:results-xray}

\begin{figure*}
   \begin{minipage}[c]{0.65\textwidth}
    \includegraphics[width=.99\textwidth]{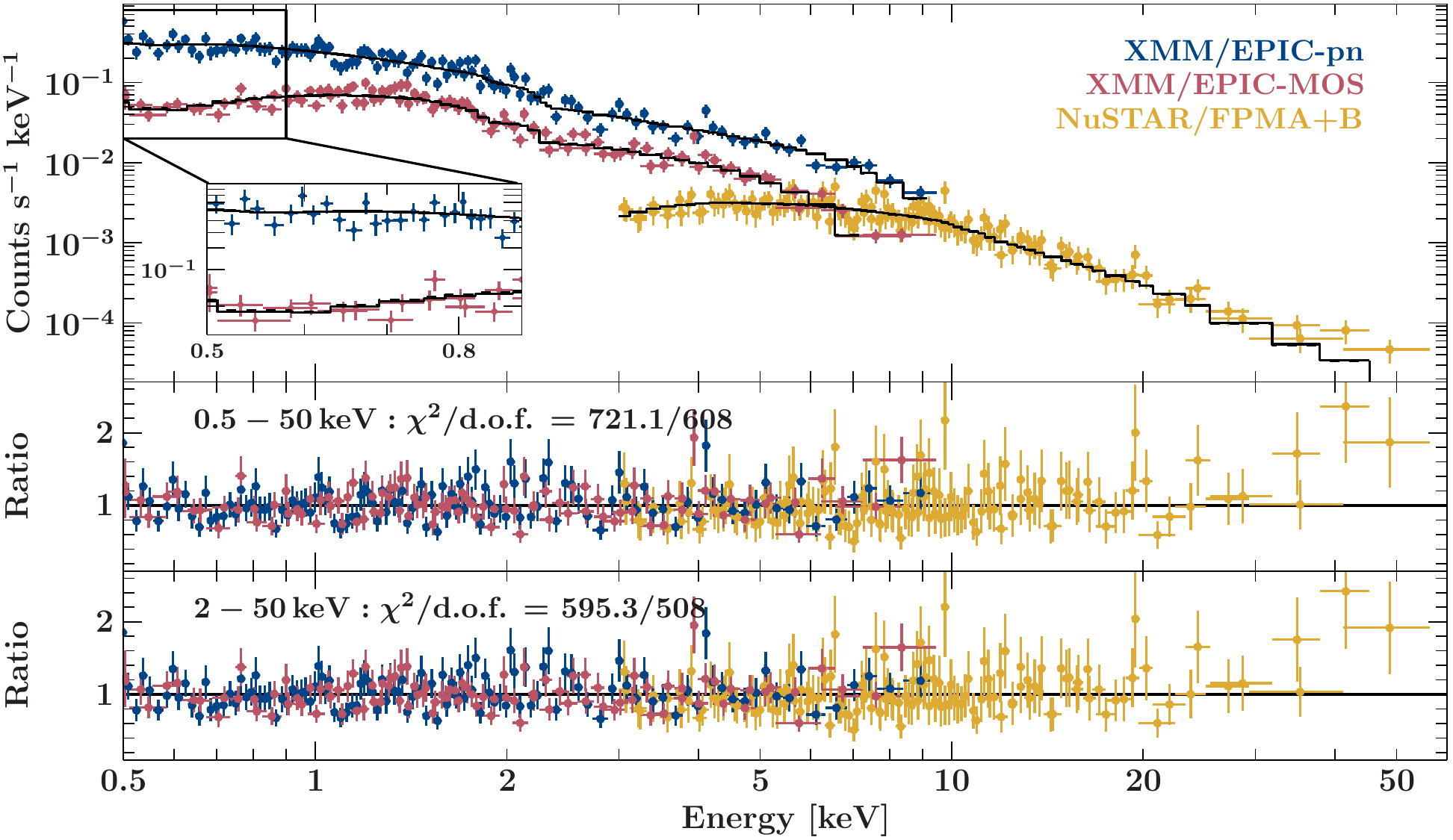}
    \end{minipage}\hfill
    \begin{minipage}[c]{0.33\textwidth}
      \caption{\textit{XMM-Newton} and \textit{NuSTAR} spectra with the best-fit results of an absorbed power-law model. For plotting purposed only the spectra are binned to a S/N of 5 and 3 for \textit{XMM-Newton} and \textit{NuSTAR}, respectively. Bins from the \textit{NuSTAR} spectrum are only shown up to 50\,keV, because at energies above this, no significant bin for the given S/N value is found.
        \textit{Upper panel:} EPIC pn (blue), MOS (red) and FPMA/FPMB (yellow) data together with the best fits for the full (0.5-80\,keV) energy range (solid line) as well as the hard (2-80\,keV) energy range (dashed line).
        The zoom window shows small deviations between the fits of the full and the hard energy range.
        \textit{Middle panel:} Data-to-model ratio for the best fit to the full energy range. \textit{Lower panel:} Data-to-model ratio for the best-fit to the hard energy range with extrapolating the model down to 0.5\,keV. }
         \label{fig:xmm_spectra}
    \end{minipage}
\end{figure*}

A feature often seen in X-ray spectra of NLSy\,1 galaxies is a soft
excess below 2\,keV \citep{vaughan1999, grupe2004}. Previously,
\citet{gallo2006} found an indication of a soft excess in PKS\,2004$-$447 in
\textit{XMM-Newton} data from 2004, while the source was in a higher
state. However, \citet{2015MNRAS.453.4037O} and
\citet{kreikenbohm2016} did not find an excess for PKS\,2004$-$447
during its low state in 2012, and \citet{kreikenbohm2016} could not
confirm the excess in the data from 2004.

Given that the source showed its brightest X-ray flux compared to 
previous observations during the \gm-ray flare reported here, we 
search for an excess below 2\,keV in the \textit{XMM-Newton} spectrum.
We apply a simple, unbroken power law model with Galactic \ion{H}{i}
absorption. 
Similar to the analysis of the individual X-ray spectra, we use C-statistics 
\citep{cash} and estimate all uncertainties at 1$\sigma$ confidence. 
All spectra are binned following the optimal binning procedure of \citet{optimalbinning}. 
Modelling both the spectra obtained with \textit{XMM-Newton} and \textit{NuSTAR} 
individually between 3 and 10 keV with a power law yields compatible values 
for the power-law indices for both instruments ($\Gamma_{3-10\,\mathrm{keV}}=1.33\pm0.09$ 
for \textit{XMM-Newton} vs. $1.37\pm0.10$ for \textit{NuSTAR}). 
Although the observations are separated by one day, this result justifies the use 
of the spectra from both instruments for a combined analysis.

First, we fitted the data in the full energy
range from 0.5\,keV to 79\,keV with a fixed $N_\mathrm{H}$, which
yields a good fit with \hbox{$\chi^2_{\mathrm{red}}=1.18$} (721.1/608) and a best-fit
power-law index of \hbox{$\Gamma=1.45\pm0.02$}. Freeing the $N_\mathrm{H}$
parameter, we find an upper limit of \hbox{$\leq0.75$} times the Galactic value
for the absorption, meaning there is no evidence for significant 
intrinsic absorption in PKS\,2004$-$447. 
Therefore, we kept this parameter fixed at the Galactic value in the further
analysis.

To search for a soft excess, we fitted the spectra again, but only for
the $2-79$\,keV band, and extrapolated the best fit down to
0.5\,keV. We show the EPIC-pn and EPIC-MOS spectra in
Fig.~\ref{fig:xmm_spectra}, where the fits to the full and the hard
energy range are shown as a solid and a dashed line, respectively.
For plotting purposes, the \textit{XMM-Newton} spectra are binned with 
a signal-to-noise (S/N) ratio of 5 per energy bin, and the \textit{NuSTAR} 
spectra with a S/N ratio of 3. 
The fit results in \hbox{$\chi^2_\mathrm{red}=1.17$} (595.3/508)
and \hbox{$\Gamma=1.45^{+0.01}_{-0.02}$}. This power-law index agrees with that
obtained from modelling the full fitted energy range. Describing the
data with a broken power law also yields no evidence for a soft
excess. We therefore conclude that there is no evidence for a soft
excess in the X-ray spectrum of PKS\,2004$-$447 during the 2019
October outburst as the fits are indistinguishable.

The presence of an iron K$\alpha$ line at 6.4\,keV is also a common
feature in NLSy\,1 galaxies. Among the small \gm-NLSy\,1 sample,
however, only 1H\,0323+342 shows an indication for an iron-line
feature. For our data, adding an unresolved Gaussian line at 6.4\,keV
does not improve the fit statistics. We determine an upper limit for
the equivalent width of $EW_{6.4\,\mathrm{keV}}\leq73$\,eV at the
90\% confidence level. This limit is slightly less constraining than
what has been reported for this source in previous analyses
\citep{gallo2006, 2015MNRAS.453.4037O, kreikenbohm2016}.

The photon index derived from the \textit{XMM-Newton} observation is 
harder compared to the values derived from the low state
analysed by \citet{2015MNRAS.453.4037O} and \citet{kreikenbohm2016},
This fits into the 'harder-when-brighter' behaviour of blazars, more precisely BL Lacs \citep[e.g.][]{giommi1990,wang2018}.
During an \textit{XMM-Newton} observation in 2004, and also at the end 
of 2013 during a monitoring campaign with \textit{Swift}, PKS\,2004$-$447
was in a bright state as well. However, a spectral hardening was
not observed at these times \citep{gallo2006, kreikenbohm2016}, 
which suggests that different processes might be responsible
for the X-ray variability that is present on monthly and yearly timescales.

\subsection{Spectral energy distribution}\label{sec:results-sed}
Due to X-ray and optical-UV observations $\sim$20\,days before, during, and
after the flare, we can construct the broadband SEDs for PKS\,2004$-$447 and
study the evolution of the SEDs over the flaring period.
We construct three SEDs, covering the time spans of a pre-flare state of the source
(MJD~58754--58770), the flaring period (MJD~58781--58784), and the
post-flare period right after the \gm-ray flare (MJD~58787--58789).
These time ranges are marked in Fig.~\ref{fig:lightcurves}.
We also consider a low activity state of this source for comparison, adopting
the period from \citet{kreikenbohm2016}.
The \textit{Fermi}-LAT data used for the low-state SED are centred on
\textit{XMM-Newton} and \textit{Swift} observations in 2012~May and March, 
respectively, and cover 24 months from 2011~May~1 through 2013~May~1.

We can describe the broadband SEDs of PKS\,2004$-$447 for all four
activity states with a simple synchrotron inverse Compton radiative
model \citep[see, e.g.][]{2009ApJ...704...38S,2009herb.book.....D,2009MNRAS.397..985G}.
In particular, the emission region is assumed to be spherical (radius
$R_\mathrm{blob}$), moving along the jet axis with the bulk Lorentz
factor $\Gamma_\mathrm{b}$. The emission region is also considered to
cover the entire cross-section of the jet whose semi-opening angle is
assumed to be 0.1~radian. We have assumed a viewing angle of 2$^{\circ}$, similar to that adopted in the modelling of beamed AGNs \citep[cf.][]{2015MNRAS.448.1060G}. Considering a significantly larger viewing angle would drastically reduce the Doppler boosting \citep[e.g.][]{1995ApJ...446L..63D}. Indeed, the observation of the large-amplitude $\gm$-ray flare from the source indicates a significant Doppler boosting as explained in the next section, hence supporting a small viewing angle.

In our model, a relativistic population of
electrons fills the emission region and emits synchrotron and inverse
Compton radiation in the presence of a uniform but tangled magnetic
field~$B$. The energy distribution of the electrons, with minimum 
and maximum energies of \gm$_{\rm min}$~and \gm$_{\rm max}$, respectively,
follows a smooth broken power law of the type

\begin{equation}
 Q(\gamma)   = Q_0\, \frac{ \gamma_\mathrm{b}^{-s_1} }{(\gamma/\gamma_\mathrm{b})^{s_1} + (\gamma/\gamma_\mathrm{b})^{s_2}},
\end{equation}
where $\gamma_\mathrm{b}$ is the break Lorentz factor, $s_1$ and $s_2$
are the spectral indices of the power law below and above
$\gamma_\mathrm{b}$, and $Q_0$ is a normalisation constant.

We include various sources of seed photons in the computation of the
inverse Compton radiation caused by the relativistic electrons present
in the jet. Sources include the synchrotron photons produced within
the emission region \citep[synchrotron self Compton or SSC, see, e.g.][]{2008ApJ...686..181F, 2019ApJ...874...47V} and thermal emission originating outside
the jet \citep[so-called external Compton or EC process; e.g.][]{1994ApJ...421..153S,2000ApJ...545..107B}.
For the latter, we
adopt radiation emitted by the accretion disk, X-ray corona, broad
line region (BLR), and the dusty torus. The comoving-frame radiative
energy densities of these AGN components have been adopted following
the prescriptions of \citet{2009MNRAS.397..985G} and are used to
derive the EC flux. Jet powers are calculated assuming a two-sided jet
and equal number density of electrons and protons, meaning we do not
consider the presence of pairs in the jet \citep[see, e.g.][]{2017MNRAS.465.3506P}. 
Protons are assumed to be cold and to
participate only in carrying the jet's momentum.

\begin{figure*}
\begin{center}
\hbox{
    \includegraphics[scale=0.6]{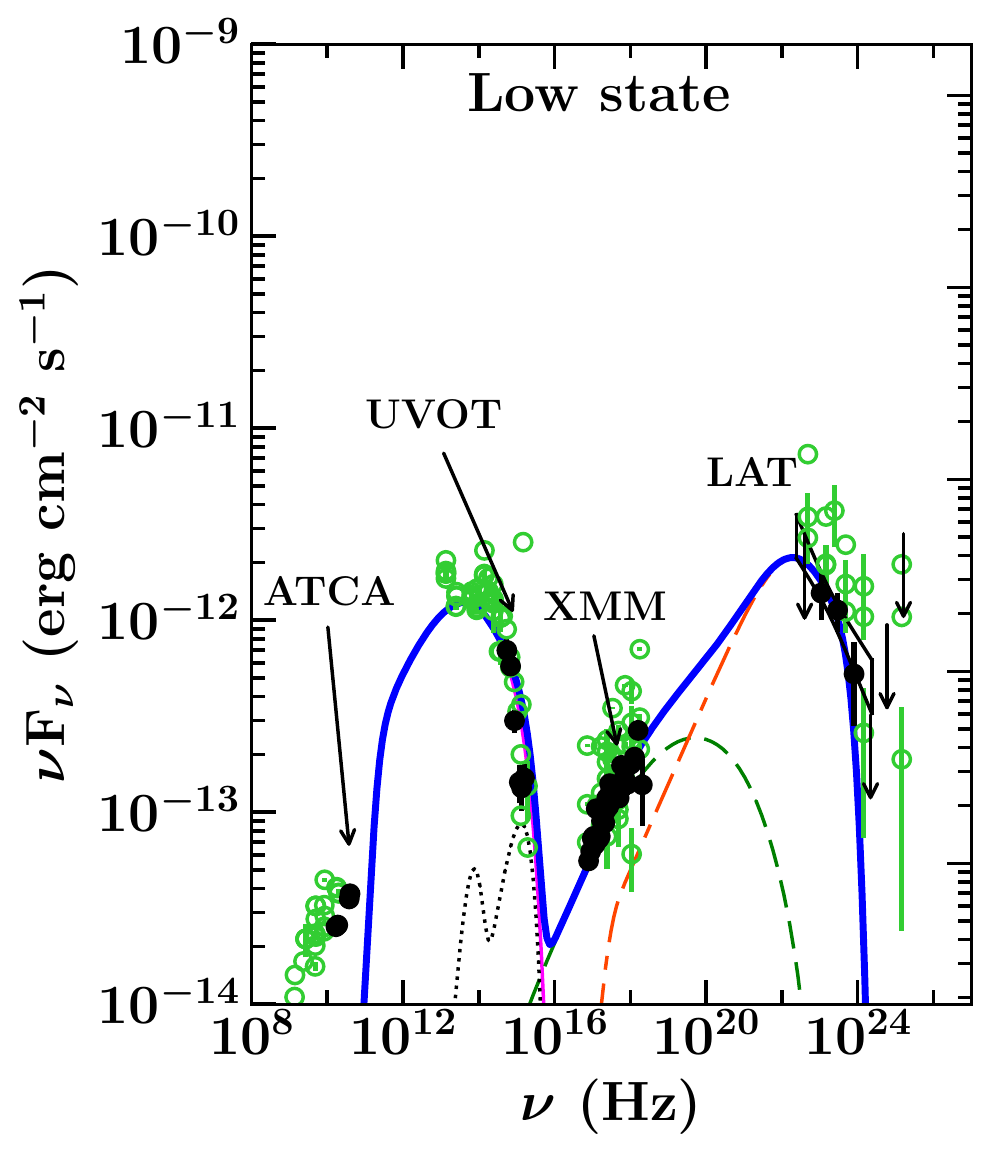}
    \includegraphics[scale=0.6]{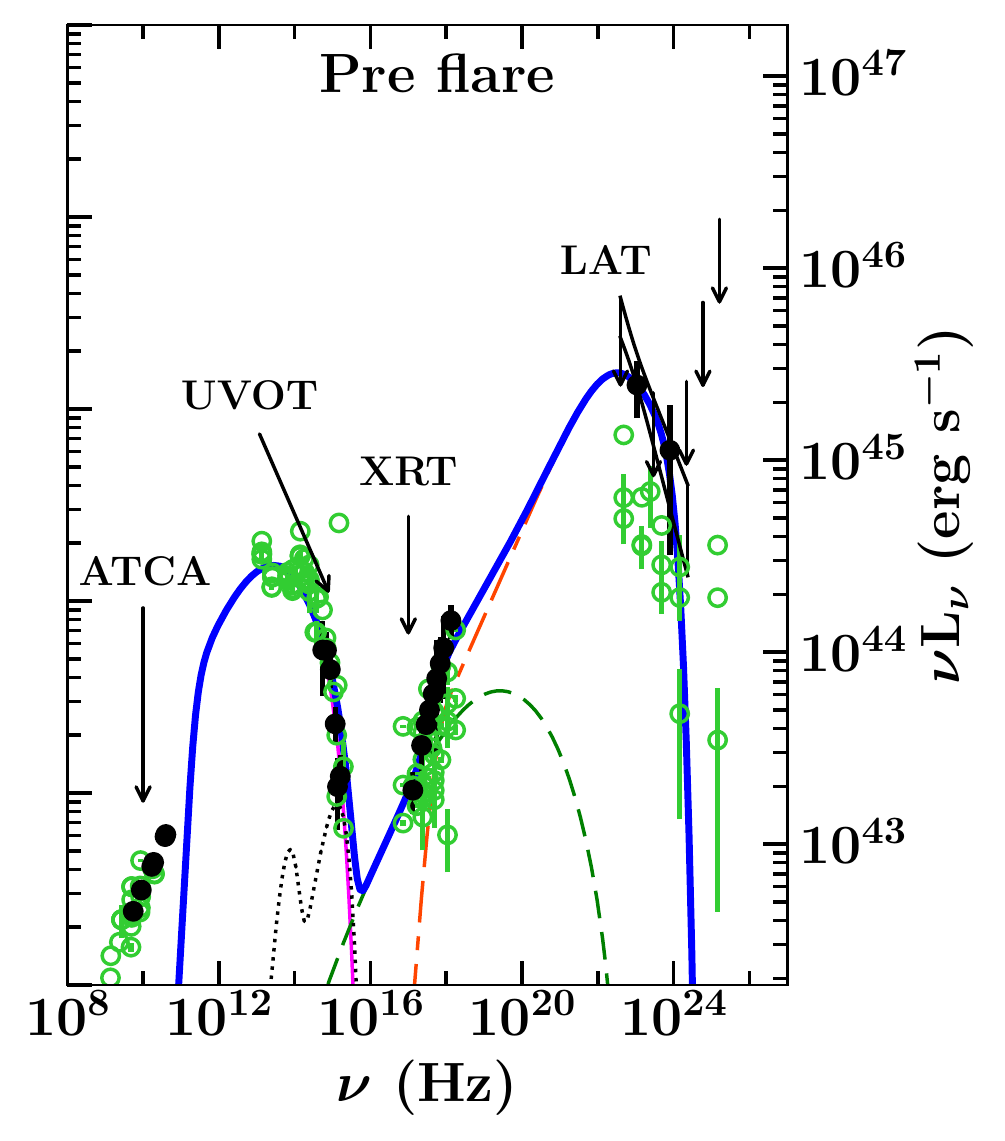}
}\hspace{0.0cm}

\hbox{
    \includegraphics[scale=0.6]{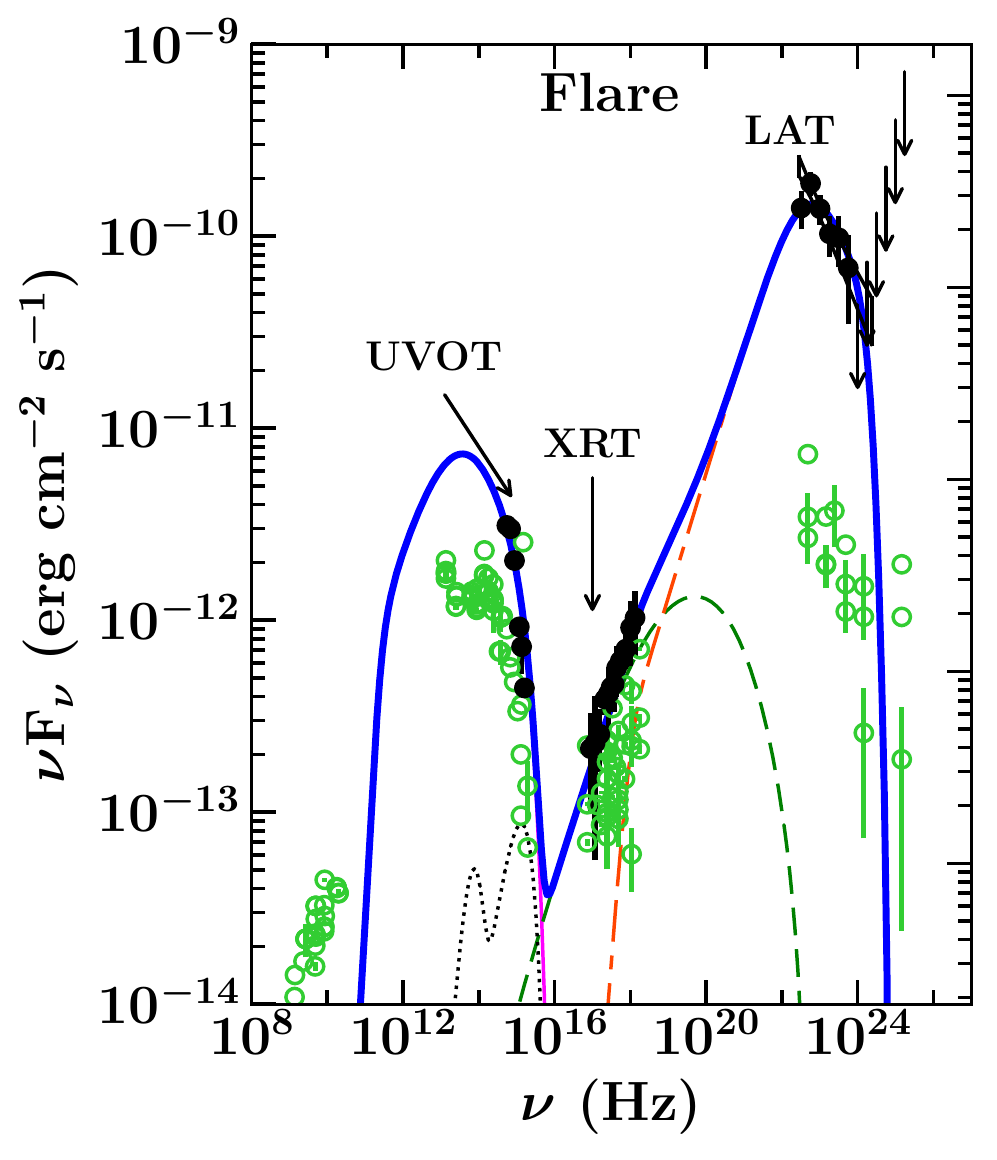}
    \includegraphics[scale=0.6]{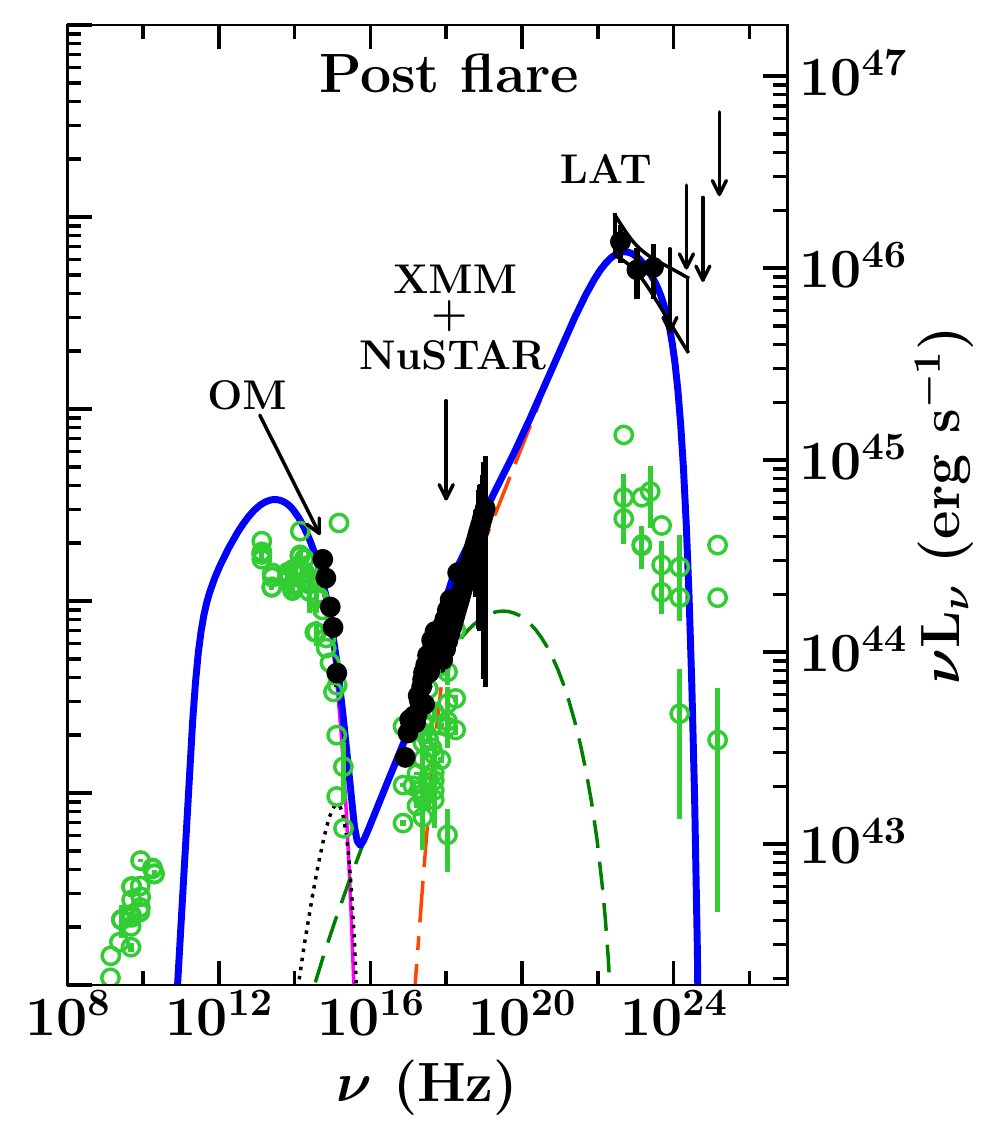}
}
        \caption{Spectral energy distribution of PKS\,2004$-$447 for four stages modelled with a simple leptonic emission model. Open green circles refer to the archival measurements taken from the SSDC SED builder (\href{https://tools.ssdc.asi.it/}{https://tools.ssdc.asi.it/}), whereas filled black circles represent the data analysed by us. The thin solid pink line shows the synchrotron emission. Dashed green and dash-dash-dot orange lines correspond to the synchrotron self Compton and external Compton processes, respectively. The dotted black line shows the thermal emission from the accretion disk and dusty torus. The thick solid blue line is the sum of all radiative components.
        The SEDs are sorted from left to right as follows: low (MJD~55682--56413), pre-flare (MJD~58754--58770), flaring (MJD~58781--58784), and post-flare activity states (MJD~58787--58789), as defined in the text. The model parameters are listed in Table~\ref{tab:sed_par}.}
        \label{fig:seds}
\end{center}
\end{figure*}

We generate the broadband SEDs following the methodology described in
Sect.~\ref{sec:datareduction} and reproduce them using the model
outlined above. The results are shown in Fig.~\ref{fig:seds} and the
associated spectral flux values and SED parameters are presented in
Tables~\ref{tab:sed_par} and~\ref{tab:sed}.
\begin{table*}[h!]
\begin{center}
\caption{Summary of the analysis of each individual energy range for each SED.\label{tab:sed_par}}
\begin{tabular}{lccccccc}
\hline\hline
 & & & \textbf{Gamma-ray}  & & & & \Tstrut\\
 & Activity state & Time bin & $\Gamma_{0.1-300~\mathrm{GeV}}$ & $F_{0.1-300~\mathrm{GeV}}$ & TS & &  \\
 &               & (MJD)  &                                & (10$^{-8}$ \phflux)  & & & \\ 
\hline
 & Low activity    & 55682$-$56413 & 2.39$\pm$0.13 & 1.2$\pm$0.3 & 50.4 & & \Tstrut\\
 & Pre-flare       & 58754$-$58770 & 2.62$\pm$0.22 & 16$\pm$4 & 38.8 &  &\\
 & Flare           & 58781$-$58784 & 2.42$\pm$0.09 & 130.0$\pm$11.6 & 472 & & \\
 & Post-flare      & 58787$-$58789 & 2.22$\pm$0.17 & 43$\pm$9 & 97 & & \\
 \hline
 & & & \textbf{Soft X-ray} & & & & \Tstrut\\
 & Activity state & Exposure & $\Gamma_{0.5-10~\mathrm{keV}}$ & Flux$_{0.5-10~\mathrm{keV}}$ & Statistics &  \\
 &                & (ksec) &                             &  & C-stat./dof  &  \\ 
\hline
 & Low activity (\textit{XMM}/pn) & 31.76 & $1.682\pm0.029$ &  $0.451\pm0.013$ & 80.36/76 & &\Tstrut \\
 & Pre-flare (XRT)    & 4.05  & $1.14\pm0.16$ & $1.53^{+0.25}_{-0.22}$ & 55.06/45 & &\Tstrut \\
 & Flare (XRT)        & 2.90  &  $1.62\pm0.18$ & $1.59^{+0.25}_{-0.22}$  & 59.59/46 & & \Tstrut\\
 & Post-flare (\textit{XMM}/pn)   & 7.77  & $1.424\pm0.024$ & $2.06\pm0.05$ &  97.23/80 &  & \Tstrut\Bstrut\\
 \hline
 & & & \textbf{Hard X-ray} & & & \Tstrut\\
 & Activity state & Exposure & $\Gamma_{3-79~\mathrm{keV}}$ & Flux$_{3-79~\mathrm{keV}}$  & Background &  Statistics &  \\
 &                & (ksec) &                             &  & norm & C-stat./dof  &  \\ 
\hline
 & Post-flare (\nustar)& 30.07 & $1.31\pm0.05$ & $6.3^{+0.5}_{-0.4}$ & $0.89\pm0.12$ (FPMA) & 357.54/331 & \Tstrut\Bstrut\\
 &     &    &    &   &  $1.07^{+0.14}_{-0.13}$  (FMPB) &    & \Tstrut\Bstrut\\
 \hline
Activity & & & \textbf{Optical-UV} &  &  &  \Tstrut\\
state & V & B & U & UVW1 & UVM2 & UVW2 &\\
  \hline
 Low  & $0.72\pm0.05$ & $0.50\pm0.03$ & $0.34\pm0.02$ & $0.18\pm0.01$ & $0.12\pm0.01$ & $0.12\pm0.01$ & \Tstrut\\
 Pre-flare    & $0.55\pm0.24$ & $0.55\pm0.14$ & $0.44\pm0.09$ & $0.23\pm0.05$ & $0.11\pm0.04$ & $0.12\pm0.04$ & \\
 Flare        & $3.12\pm0.24$ & $2.99\pm0.14$ & $2.05\pm0.10$ & $0.92\pm0.06$ & $0.73\pm0.20$ & $0.44\pm0.03$ & \\
 Post-flare   & $2.8\pm0.3$ & $1.90\pm0.18$ & $1.32\pm0.14$ & $0.65\pm0.11$ & $0.31\pm0.07$ & $0.27\pm0.05$ & \\
  \hline
\end{tabular}
\end{center}
\tablefoot{The X-ray and optical-UV fluxes are in units of 10$^{-12}$ \ergflux. For the X-rays, we report the unabsorbed flux.}
\end{table*}

\begin{table*}
{\small
\begin{center}
\caption{Summary of the parameters used for and derived from the modelling of the multi-epoch SEDs of PKS\,2004$-$447 shown in Fig.~\ref{fig:seds}. The central black hole mass and the accretion disk luminosity are taken as $7\times10^7$ \msun~and $2\times10^{43}$ \lum, respectively, and we assume the characteristic temperature of the IR-torus to be 1100 K. A viewing angle of 2$^{\circ}$ is adopted. For the given accretion disk luminosity, the size of the BLR and dusty torus are 4.6$\times$10$^{-3}$~pc and 3.4$\times$10$^{-2}$~pc, respectively. We note that the jet powers are computed by assuming a two-sided jet.}\label{tab:sed}
\begin{tabular}{lccccc}
\hline
Parameter                                            &  Symbol  & Low Activity & Pre-flare & Flare & Post-flare \\
\hline
Particle spectral index before break energy & $s_1$                       & 2.1        & 2.1  & 1.7  & 2.0   \\
Particle spectral index after break energy   & $s_2$                      & 4.0        & 4.0  & 4.0  & 4.0   \\
Minimum Lorentz factor of the particle distribution  & $\gamma'_\mathrm{min}$  & 4           & 4    & 4    & 4  \\
Break Lorentz factor of the particle distribution    & $\gamma'_\mathrm{b}$    & $1.3\times10^{3}$ &$1.1\times10^{3}$ & $0.9\times10^{3}$ &  $1.1\times10^{3}$ \\
Maximum Lorentz factor of the particle distribution  & $\gamma'_\mathrm{max}$  & 6000       & 5500 & 5000 & 5000 \\
Particle energy density, in erg cm$^{-3}$            & $U'_\mathrm{e}$         & 0.18       & 0.22 & 0.10 & 0.22 \\
Magnetic field, in Gauss                             & {$B$}                   & 0.4        & 0.3  & 0.3  & 0.3  \\
Bulk Lorentz factor                                  &$\Gamma_\mathrm{b}$      & 11         & 20   & 26   & 24   \\
Dissipation distance, in $10^{-2}$ parsec            & $R_\mathrm{dist}$       & 2.01       & 2.01 & 2.01 & 2.01  \\
Size of the emission region, in $10^{15}$cm          & $R_\mathrm{blob}$       & 6.2       & 6.2 & 6.2 & 5.17     \\
Compton dominance                                    & $CD$                 & 2          & 10   & 18   & 20     \\
\hline
Jet power in electrons, in \lum, in log scale        & $P_\mathrm{e}$      & 44.2  & 44.8 & 44.7 & 45.0 \\
Jet power in magnetic field, in \lum, in log scale   & $P_\mathrm{B}$      & 42.7  & 43.0 & 43.2 & 43.0 \\
Radiative jet power, in \lum, in log scale           & $P_\mathrm{r}$      & 43.7  & 44.9 & 45.2 & 45.2\\
Jet power in protons, in \lum, in log scale          & $P_\mathrm{p}$      & 46.2  & 46.8 & 46.3 & 46.9 \\
\hline
\end{tabular}
\end{center}
}
\end{table*}

\section{Discussion}\label{sec:discussion}

\subsection{Variability}
The detection of a \gm-ray flare from PKS\,2004$-$447 provides more observational evidence supporting the blazar-like behaviour of \gm-NLSy~1 galaxies \citep[see also e.g.][]{2016MNRAS.458L..69B}.
In general, \gm-ray variability is an indicator for the presence of a closely aligned,
relativistic jet. 
For blazars variability on timescales as short has minutes is commonly observed 
at TeV energies \citep[e.g.][]{rieger2010, aleksic2011}, but such 
short timescales have only been observed in few sources at GeV energies  \citep[e.g.][]{meyer2019}.

It has been proposed that \gm-NLSy\,1s represent the start of the life
of an AGN, when the central black hole mass is still below $10^8\msun$
and the source appears not as bright as a full-grown FSRQ
\citep{marthur2000, foschini2017, reviewNLSy1_paliya2019}. Nevertheless, the detection of blazar-like short-term variability in \gm-NLSy\,1
galaxies, which has been seen previously in 1H\,0323+342 \citep[flux doubling timescales of $\sim3$h;][]{paliya2014} and PKS\,1502+036 \citep[variability on 12\,h timescales;][]{dammando_PKS1502}, suggests that the physical mechanisms operating in the relativistic jet of \gm-NLSy~1s are similar to those working in blazar jets resulting in fast $\gamma$-ray variability \citep[e.g.][]{shukla2020}.
For PKS\,2004$-$447, we find indications for sub-daily variability with
flux doubling times as short as 2.2~hours at a $2.8\sigma$~level.

We note that if we consider only GeV flares with fluxes above
$10^{-6}\,\mathrm{ph}\,\mathrm{cm}^{-2}\,\mathrm{s}^{-1}$, the few
observed flares by \gm-NLSy\,1 galaxies lasted roughly 1--4\,days.
Blazar flares reported from single sources at \gm-ray energies have a tendency to be
brighter, last longer, and for some sources occur more often, but a
strong bias towards the most luminous and extreme detections exists due
to the different sensitivities and observing constraints of space-based and
ground-based \gm-ray telescopes:
While the large field of view and observing strategy of \textit{Fermi}-LAT 
offers unbiased all-sky observations of many blazars in 
the GeV energy regime, its relatively small collection area renders it less 
sensitive to weak flares. In the TeV energy range Cherenkov
telescopes have large collection areas giving them good sensitivity to short time
variability, but with their relatively small fields of view and low duty cycles,
their observations are limited to a smaller sub-sample of targeted observations on blazars. 
Even though observing programmes often include scheduled observations on a selection
of blazars during the parts of the year that they are visible from the ground, many
blazar observations are triggered and therefore take place during a particularly active period.
This leads to an under-reporting of short, less luminous blazar
flares, which could in turn belong to a class of less luminous blazars that
might be missing in the AGN evolution scenario.

\subsection{Physical interpretation of the SED parameters}
We now turn to the interpretation of the SED of PKS\,2004$-$447. We
first discuss the low activity state SED of the source before
considering the \gm-ray flaring epochs. The motivation is to first
derive the quiescent-level SED parameters and then to explain the
flaring-state SEDs while changing the minimum number of input
parameters. This approach may permit us to understand the primary
factors responsible for the \gm-ray flare.

We have used the H$\beta$ emission line dispersion (1869 km~s$^{-1}$) and luminosity ($1.62\times10^{42}$ \lum) reported in \citet[][]{2017A&A...603C...1F}, which were calculated from the optical spectrum of PKS 2004$-$447 published in \citet[][]{1997MNRAS.284...85D} to derive a value of the mass of the black hole of $M_\mathrm{BH}=7\times10^7 M_\odot$. This value is similar to that usually determined for radio-loud NLSy~1 galaxies \citep[][]{2017A&A...603C...1F} but lower than typical blazars \citep[e.g.][]{2012ApJ...748...49S}. 
Interestingly, \citet[][]{2016MNRAS.458L..69B} have reported a $M_{\rm BH}$ of $6\times10^8~M_{\odot}$ from the spectropolarimetric observation of PKS 2004$-$447. Furthermore, the accretion disk luminosity reported in \citet[][$L_{\rm disk}=4.8 \times 10^{44}$ \lum]{2017A&A...603C...1F} is not supported by the shape of the low activity state optical-UV spectrum of PKS 2004$-$447.
This is because the radiation from such a luminous accretion disk should be detectable in the form of a strong big blue bump at optical-UV frequencies, especially during the low jet activity state. As can be seen in the top left panel of Fig.~\ref{fig:seds}, no such feature was observed. Rather, by reproducing the observed optical-UV emission with the combined synchrotron and accretion disk spectra, we have constrained the $L_{\rm disk}$ value to $2\times10^{43}$ \lum. A larger $L_{\rm disk}$ value would be in disagreement with the data. For the adopted values of $M_{\rm BH}$ and $L_{\rm disk}$, the accretion rate in Eddington units is $\sim$0.4\% of the Eddington one. This is lower than $\sim$1\%, which is typically found for radiatively efficient accreting systems, that is AGNs exhibiting broad emission lines \citep[e.g.][]{2017MNRAS.469..255G,2019ApJ...872..169P}. This observation indicates that PKS\,2004$-$447 could be one of the rare AGNs where the accretion process can be radiatively efficient even with such a low level of accretion activity \citep[see, e.g. Fig.~6 of][]{2012MNRAS.421.1569B}.

For a look at the evolution of the SED we start with the low activity state.
There, the infrared-to-ultraviolet (IR-to-UV) spectrum of PKS\,2004$-$447
is well explained by the synchrotron emission.
We note that the emission from a giant elliptical galaxy could explain part of 
the observed IR-optical SED. Though the wavelength at which the host
galaxy emission peaks ($\sim$1.5 micron) is not covered in our multi-band 
follow-up campaign, the observed flux variability and the overall shape 
of the optical-UV spectrum (Fig.~\ref{fig:lightcurves} and Fig.~\ref{fig:seds})
point to the synchrotron origin of the observed emission.

The shape of the steeply falling optical-to-UV SED enables us to
constrain the high-energy spectral index of the particle spectrum,
$s_2$, and also the break, $\gamma_\mathrm{b}$, and maximum energy of
the emitting electron population, $\gamma_\mathrm{max}$. On the other
hand, we use the shape of the X-ray spectrum to constrain the
low-energy particle spectral index, $s_1$. The model fails to explain
the low-frequency radio data because compact emission regions are
synchrotron-self absorbed at low radio frequencies. Additional
emission components on larger scales (approaching parsec scales and
beyond) dominate the observed radio emission, but are not included in
the model.

In the low activity state, the X-ray emission is mainly due to SSC.
The level of SSC along with the constrained synchrotron emission
allows us to derive the size of the emission region and the magnetic
field (see Table \ref{tab:sed_par}).
The \gm-ray emission is explained with the external Compton process
with seed photons provided by the dusty torus. This sets the location
of the emission region outside the BLR but inside the dusty torus. The
shape of the \gm-ray spectrum provides further constraints to $s_2$.
Overall, the derived SED parameters during the low activity state are
similar to previous studies
\citep{2013ApJ...768...52P,2015MNRAS.453.4037O}. There are a few minor
differences, for example the large bulk Lorentz factor reported by
\citet{2015MNRAS.453.4037O}, which could be mainly due to datasets
taken at different epochs and underlying assumptions associated with
the adopted leptonic models. In particular, and in contrast to what we do here,
the X-ray part of the SED during a different low activity state of the source 
was modelled by \citet{2015MNRAS.453.4037O} with the EC of
seed photons from the dusty torus.

During the pre-flare phase, \swift-UVOT shows that the level of the
optical-UV emission, and hence the synchrotron radiation, remains
comparable to that measured during the quiescent state
(Fig.~\ref{fig:seds}). In addition to the X-ray flux increase, a significant spectral
hardening is observed in the 0.5--10\,keV energy
range (see Table~\ref{tab:sed_par}). We explain the X-ray spectrum
with a combination of the SSC and EC processes. There is a slight
decrease in the magnetic field (from 0.4\,G to 0.3\,G) causing an
enhancement of both SSC and EC fluxes since the level of the
synchrotron emission remains the same. This is because a decrease in
the magnetic field strength requires an increase in the number of
electrons to produce the same level of the synchrotron flux, which
ultimately enhances the inverse Compton flux
\citep[e.g.][]{2009ApJ...692...32D}. Above a few keV, the modelled
pre-flare SED includes a significant contribution from EC photons. The
different electron populations of SSC and EC producing electrons is
responsible for the flattening of the X-ray spectrum. Furthermore, the
increase in the \gm-ray flux is even larger compared to that seen
in the X-ray band. The ratio of the inverse Compton to synchrotron
peak luminosities, meaning Compton dominance \citep[see,
e.g.,][]{2013ApJ...763..134F} significantly increases with respect to
the low activity state. We explain this observation with an increase
in the bulk Lorentz factor and, hence, a larger Doppler boosting.

At the peak of the flare, the enhancement of the \gm-ray flux is
largest with respect to that observed at lower frequencies
(Fig.~\ref{fig:seds}). There is an increase in the optical-UV flux
level indicating an elevated electron energy, possibly due to a fresh
supply of energetic electrons within the emission region. Accordingly,
the synchrotron energy density gets enhanced, leading to the
brightening of the SSC radiation, as observed in the X-ray band.
However, the larger amplitude of the flux variations seen at
\gm-rays indicates an even stronger boosting. Since we explain the \gm-ray
spectrum with the EC process, this can be understood as follows: In
addition to the usual beaming due to the relativistic motion of the plasma
towards the observer, the external photon field receives an additional
boosting due to motion of the emission region with respect to the
photon fields external to the jet \citep[see][]{1995ApJ...446L..63D}.
This phenomenon causes the radiation pattern of the external Compton
emission to be anisotropic even in the comoving frame, making it more
sensitive to Doppler boosting with respect to the synchrotron and SSC
mechanisms. The relatively large enhancement of the \gm-ray flux could
therefore be due to increased Doppler boosting. This conjecture is
confirmed by the SED modelling, where we find the bulk Lorentz factor
to increase considerably during the flare. This also makes the EC peak
much more luminous than the synchrotron peak, thus indicating a
Compton dominated SED, which is observed (see Fig.~\ref{fig:seds}).

The data from PKS\,2004$-$447 taken after the peak of the \gm-ray flare
reveal a decrease in the optical-UV and X-ray fluxes, similar to that
found during the pre-flare state, while the source remains bright in
the $\gamma$-rays, indicating a still large Doppler beaming, albeit a
bit lower than that at the flare peak (see Table~\ref{tab:sed}). Comparing the
photon indices derived from the individual \xmm~and \nustar~data 
during the post-flare period (see Table \ref{tab:sed_par}, a slight
spectral hardening is visible at hard X-ray energies, even when we consider the
influence of the background. As discussed
above, this observation implies a dominance of the SSC mechanism at
softer X-rays ($<$10\,keV). At higher energies, the EC process takes over, leading
to the observation of the flatter \nustar~spectrum. These findings
reveal the crucial role of data taken above 10\,keV in disentangling
various radiative mechanisms at work during the flare.

When we compared various jet powers estimated during different activity
states, an interesting pattern was noticed. Compared to the
low state, the jet powers increase during the pre-flare phase,
particularly the radiative and kinetic powers, $P_\mathrm{r}$ and
$P_\mathrm{p}$, in which a change of about one order of magnitude more is
found. During the main flare, there is a significant enhancement of $P_\mathrm{r}$; however, the kinetic luminosity decreases. This behaviour suggests an efficient
conversion of $P_\mathrm{p}$ into radiation during
the flare since $P_\mathrm{p}$ has increased in the post-flare state, again. This presence of
radiatively efficient jets during \gm-ray flares has also been
reported for blazars \citep[see, e.g.][]{2011ApJ...733...19T,2013ApJ...766L..11S,2015ApJ...803...15P}.
Interestingly, the observed radiative power
exceeds the total available accretion power. Given the fact
that this is the first GeV outburst detected from PKS\,2004$-$447 in
more than a decade of \fermi-LAT operation, such extraordinary events
can be considered rather rare and short-lived
\citep[e.g.][]{2010MNRAS.405L..94T}.

\begin{figure}
   \centering
    \includegraphics[width=\linewidth]{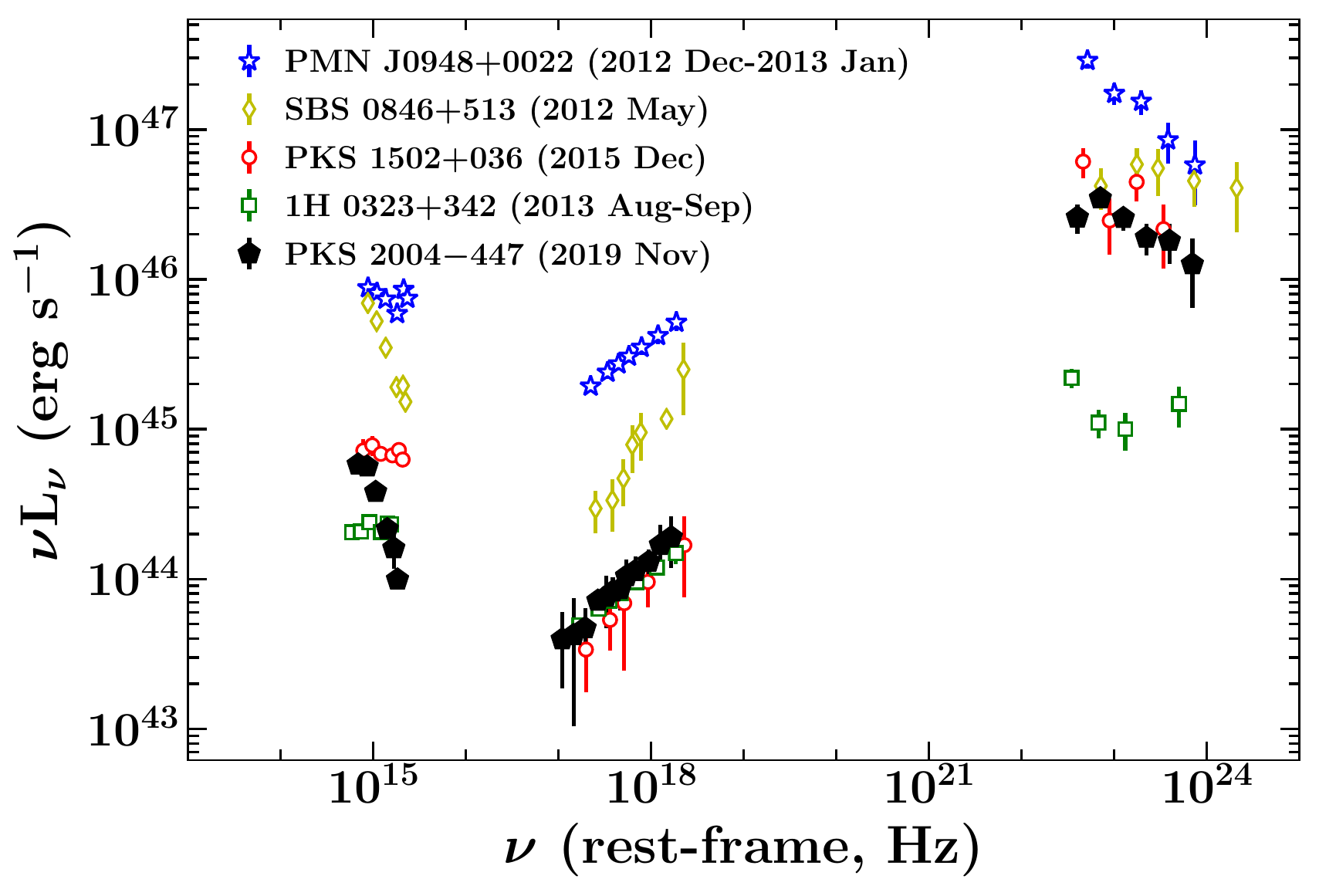}
     \caption{Comparison of the rest-frame SEDs of \gm-ray flaring NLSy\,1 galaxies covering their GeV flaring epochs. The epochs of the flares are given in their labels. Except for PKS\,2004$-$447, data taken for this comparison were analysed by \citet{paliya2016} and \citet{2013MNRAS.436..191D}.}
    \label{fig:sed_comp}
\end{figure}

\subsection{Comparison with flares from other \gm-NLSy\,1 galaxies}
Since the sample of flaring \gm-NLSy\,1 galaxies is so small, not much
is known about their typical flaring behaviour nor about their possible
differences. We compare the flaring state broadband SED of
PKS\,2004$-$447 with four \gm-NLSy\,1 galaxies that have shown a GeV
flare in the past. The data for 1H\,0323+342, PMN\,J0948+0022, and
PKS\,1502+036 are taken from \citet{paliya2016}. Although the
brightest flare of SBS\,0846+513, which happened in 2011, is not
covered by multi-wavelength data, this source showed high \gm-ray
activity in May 2012 as well. During that time, the \gm-ray emission
was slightly below the flare of 2011, but X-ray and optical-UV data
are available. We take data of SBS\,0846+513 from
\citet{2013MNRAS.436..191D}. All SEDs obtained from data throughout
their respective flares are plotted in Fig.~\ref{fig:sed_comp}.
The parameters discussed in this comparison are also listed in 
Table~\ref{tab:comparison_nlsys}.

\begin{table*}[]
\caption{Comparison of chosen SED model parameters of the flaring \gm-NLSy\,1 galaxies. The abbreviations for the papers refer to the following papers: G-2020: this paper; P-2014: \citet{paliya2014}; P\&S-2016: \citet{paliya2016}; D'A-2013: \citet{2013MNRAS.436..191D}; D'A-2015:  \citet{dammando2015}}\label{tab:comparison_nlsys}
{\small
\begin{center}
\begin{tabular}{lccccc}
\hline\hline
  & PKS\,2004$-$447 & 1H\,0323+342 & PKS\,1502+036  & SBS\,0846+513   & PMN\,J0948+0022 \Tstrut\\ 
Reference    & G-2020  & P-2014 & P\&S-2016 & D'A-2013 & D'A-2015 \\ 
\hline
Redshift, $z$ & 0.24  & 0.061 & 0.409 & 0.5835  & 0.584 \Tstrut\\ 
\hline
Optical/UV: &  &  &  &  &  \Tstrut\\
L$_{\mathrm{disk}}$ [erg s$^{-1}$]  & $2\times10^{43}$ & $1\times10^{45}$ & $6\times10^{44}$ & $4.4\times10^{43}$ & $5.7\times10^{45}$ \\
Origin of emission            & Synchrotron & Accretion disk & Synchrotron + & Synchrotron & Synchrotron + \\ 
                &  &  & accretion disk & & accretion disk \\
\hline
X-ray:     &  &  &  &  & \Tstrut\\
Index                         & $1.38\pm0.29$ & $1.55\pm0.08$ & $1.33\pm0.56$ & $1.6\pm0.3$ & $1.55\pm0.11$ \\
Dominance in soft X-rays      & SSC & EC & EC & EC & EC \\
\hline
\gm-ray:    &  &  &  &  &  \Tstrut\\
$\Gamma_{0.1-300\,\mathrm{GeV}}$  & $2.42\pm0.09$ & $2.47\pm0.11$ & $2.57\pm0.16$ & $2.13\pm0.05$ & $2.65\pm0.11$ \\
L$_{\gamma}$ [erg s$^{-1}$]    & $2.9\times10^{47}$ & $4.7\times10^{46}$ & $1.2\times10^{48}$ & $5\times10^{47}-10^{48}$ & $1.5\times10^{48}$ \\
$\Gamma_\mathrm{b}$            & 26 & 7 & 25 & 40 & 30 \\
Compton process                & EC/Torus & EC/BLR & EC/Torus & EC/Torus & EC/Torus \\
\hline
\end{tabular}
\end{center}
}
\end{table*}

The shape of the optical-UV emission from PKS\,2004$-$447 is similar
to that of SBS\,0846+513, but differs from the other three sources:
Although the level of luminosity is different for 1H\,0323+342,
PMN\,J0948+0022, and PKS\,1502+036, their observed optical-UV spectra
could all be explained with a combination of the synchrotron and
accretion disk emission \citep{paliya2014}. For 1H\,0323+342, the
optical-UV emission remains disk dominated even during the GeV flare.
Such thermal emission, however, is not observed in PKS\,2004$-$447
(see also Fig.~\ref{fig:seds}). Taken together with SED modelling, this
indicates that emission from the accretion disk is negligible compared
to the synchrotron emission.

In the X-rays, the luminosity of PKS\,2004$-$447 is similar to that of
PKS\,1502+036 and 1H\,0323+342. Within the soft X-rays, models predict
a transition from SSC to EC emission. During flaring states, the EC
can become dominant over SSC, as it is the case for 1H\,0323+342,
PMN\,J0948+0022 and PKS\,1502+036. The EC component of
PKS\,2004$-$447, however, starts at higher energies compared to the
other \gm-NLSy\,1 galaxies; therefore most of its soft X-ray emission
originates from SSC. The spectral shapes of \gm-NLSy\,1 galaxies
behave similarly by showing a harder spectral index during a flare
compared to low states. All sources, including PKS\,2004$-$447, show
photon indices of $\sim$1.3-1.6.

In the $\gamma$-rays, PKS\,2004$-$447 reaches about the same
luminosity as PKS\,1502+036, and is about one order of magnitude more luminous than
1H\,0323+342 (see Fig.~\ref{fig:sed_comp}). SBS\,0846+513 showed a
slightly higher luminosity, while PMN\,J0948+0022 presents the highest
luminosity ever observed for a \gm-ray flare of a NLSy\,1 galaxy,
exceeding the luminosity of PKS\,2004-447 by a factor of ten. The
\gm-ray photon indices during the flaring state are ${\sim}2.5$, with
the exception of SBS\,0846+513, which shows a significantly harder
spectrum. 
Both PKS\,2004$-$447 and PKS\,1502+036 exhibit a
similar luminosity and bulk Lorentz factor \citep[PKS\,1502+036
has $\Gamma=25$;][]{paliya2016}. The mass of their central black holes
is similar as well. However, the accretion disk luminosity of the latter is a factor
of $\sim$30 larger indicating a higher accretion rate in Eddington units.
Since the sizes of the BLR and the torus adjust to the luminosity of the
accretion disk, they are likely ten times larger as well for
PKS\,1502+036. For PKS\,2004$-$447, in order to be able to produce a
similar \gm-ray luminosity, a higher particle density is required.

The $\gamma$-ray emission of PKS\,2004$-$447 is explained by the EC
process with seed photons provided by the dusty torus, as already
reported for SBS\,0846$+$513, PMN\,J0948$+$0022, and PKS\,1502$+$036
during flaring episodes
\citep{2013MNRAS.436..191D,dammando2015,dammando_PKS1502}. As is observed
for PKS\,2004$-$447, a high Compton dominance has also been seen in
these sources at the peak of the activity, confirming that the EC
emission is the main mechanism for producing $\gamma$-rays, similar to
several FSRQs. This result confirms the similarities between
$\gamma$-NLSy\,1s and FSRQs. In contrast to PKS\,2004$-$447, for which
an increase in the bulk Lorentz factor is the driver of the change in
the SED for different activity states, comparing low and flaring
activity states, the SEDs of SBS\,0846$+$513, PMN\,J0948$+$0022, and
PKS\,1502$+$036 can be described satisfactorily by changing the
electron distribution parameters as well as the magnetic field. In the
same way, a significant shift of the synchrotron peak has been
observed during the flaring states of SBS\,0846$+$513 and
PMN\,J0948$+$0022, while it was not observed for PKS\,2004$-$447.

\section{Summary}\label{sec:conclusion}
In this paper we presented the analysis of the first GeV flare from the
\gm-NLSy\,1 galaxy PKS\,2004$-$447. We created daily and sub-daily
binned \textit{Fermi}-LAT light curves and used the method of
\citet{meyer2019} to derive the temporal behaviour of the flare. We
built SEDs containing (quasi-)simultaneous datasets, and modelled the
emission of the source during a low state, as well as before, during
and after the flare, with a one-zone, leptonic model. Our main results
are the following:
\begin{enumerate}
\item Short-term variability on timescales of hours to weeks is found
  in the \gm-rays. The X-ray and optical-UV light curves show flux changes within days. The ATCA light curves show a steady rise since 2018. Observations before
  and after the flare revealed a slight increase, which seemed to be
  strongest at 5.5\,GHz. We find indications for $\gamma$-ray flux-doubling
  times as short as $\sim2.2$\,hours at the ${\sim}3\sigma$~level.
\item The soft excess frequently observed in the X-ray spectra of NLSy\,1
  galaxies is not found in the \textit{XMM-Newton} spectrum of
  PKS\,2004$-$447 during the flare, which is in agreement with the 2012
  low-state analysis of \citet{kreikenbohm2016}.
\item The simultaneous multi-wavelength data can be well described
  with a one-zone leptonic model. The emission region
  from where the flare possibly originated lies outside the BLR, but
  within the dusty torus. The \gm-ray emission is dominated by EC
  processes, while the SEDs of the flaring, pre-, and post-flare state
  exhibit strong Compton dominance. During the GeV flare, the source
  was in an elevated activity state at lower frequencies as well.
  However, the amplitude of the flux variability was the highest in
  the \gm-ray band. SED modelling explains this behaviour with an
  increase in the bulk Lorentz factor of the jet, similar to that
  typically seen in powerful FSRQs.
\end{enumerate}
We conclude that all of the observations of PKS\,2004$-$447, and \gm-NLSy1 in general, point to a scenario in which these objects could be considered to belong to the blazar sub-class of radio-loud emitters.
\begin{acknowledgements}
 We are grateful to the journal referee for a constructive criticism. We thank the \textit{Fermi}-LAT collaboration members E.~Ros, D.~Horan, D.~J.~Thompson, and G.~Johannesson for their comments, which helped to improve the manuscript.
  A. Gokus was partially funded by the Bundesministerium für
  Wirtschaft und Technologie under Deutsches Zentrum für Luft- und
  Raumfahrt (DLR grant number 50OR1607O) and by the German Science
  Foundation (DFG grant number KR 3338/4-1).
  V.S.P.'s work was supported by the Initiative and Networking Fund of
  the Helmholtz Association.
  S.M.W. acknowledges support by the Stiftung der deutschen Wirtschaft (sdw).
  F.D.\ acknowledges financial contribution from the agreement
  ASI-INAF n. 2017-14-H.0.
      
  We are grateful to the ATCA, \textit{Swift}, \textit{XMM-Newton},
  and \textit{NuSTAR} PIs for approving the ToO observations, and to
  the mission operation teams for quickly executing them.
      
  The \textit{Fermi} LAT Collaboration acknowledges generous ongoing
  support from a number of agencies and institutes that have supported
  both the development and the operation of the LAT as well as
  scientific data analysis. These include the National Aeronautics and
  Space Administration and the Department of Energy in the United
  States, the Commissariat \`a l'Energie Atomique and the Centre
  National de la Recherche Scientifique / Institut National de
  Physique Nucl\'eaire et de Physique des Particules in France, the
  Agenzia Spaziale Italiana and the Istituto Nazionale di Fisica
  Nucleare in Italy, the Ministry of Education, Culture, Sports,
  Science and Technology (MEXT), High Energy Accelerator Research
  Organization (KEK) and Japan Aerospace Exploration Agency (JAXA) in
  Japan, and the K.A.~Wallenberg Foundation, the Swedish Research
  Council and the Swedish National Space Board in Sweden. Additional
  support for science analysis during the operations phase is
  gratefully acknowledged from the Istituto Nazionale di Astrofisica
  in Italy and the Centre National d'\'Etudes Spatiales in France.
  This work was performed in part under DOE Contract
  DE-AC02-76SF00515.
      
      This work made use of data from the \textit{NuSTAR} mission, a project led by the California Institute of Technology, managed by the Jet Propulsion Laboratory, and funded by the National Aeronautics and Space Administration. We thank the \textit{NuSTAR} Operations, Software, and Calibration teams for support with the execution and analysis of these observations. This research has made use of the \textit{NuSTAR} Data Analysis Software (NuSTARDAS) jointly developed by the ASI Science Data Center (ASDC, Italy) and the California Institute of Technology (USA).
      
      We acknowledge the use of public data from the Swift data archive.
      
      The Australia Telescope Compact Array is part of the Australia Telescope National Facility which is funded by the Australian Government for operation as a National Facility managed by CSIRO.
      
      This research has made use of a collection of ISIS functions (ISISscripts) provided by ECAP/Remeis
      observatory and MIT (http://www.sternwarte.uni-erlangen.de/isis/).
      
      The colours in Fig.~\ref{fig:lightcurves}, Fig.~\ref{fig:diffbin}, Fig.~\ref{fig:atca_lc}, and Fig.~\ref{fig:xmm_spectra} were taken from Paul Tol's colour schemes and templates (https://personal.sron.nl/~pault/).

\end{acknowledgements}

\bibliographystyle{aa} 
\bibliography{pks2004-447} 
\begin{appendix}
\section{Flux densities measured with ATCA}
The radio data in Table \ref{tab:atca_data} cover ATCA observations from
2010 up to early 2020 of PKS\,2004$-$447 and are taken in three different 
bands ( $\lambda$4-cm, $\lambda$15-mm, and $\lambda$7-mm).
\begin{table*}[h]
    \caption{ATCA flux densities taken in the $\lambda$4-cm (5.5~GHz,9~GHz), $\lambda$15-mm (16.8~GHz, 17~GHz, 19~GHz, 21.2~GHz), and $\lambda$7-mm band (38~GHz, 40~GHz). Fluxes are given in mJy, and only statistical uncertainties are reported. Remarks: $^{\dagger}$Observation done at 16.8~GHz.  $^{\ddagger}$Observation done at 21.2~GHz}
    \centering
    \label{tab:atca_data}
        \begin{tabular}{lcccccc}
        \hline\hline
        MJD & S$_{\mathrm{5.5~GHz}}$ & S$_{\mathrm{9~GHz}}$ & S$_{\mathrm{17~GHz}}$ & S$_{\mathrm{19~GHz}}$ & S$_{\mathrm{38~GHz}}$ & S$_{\mathrm{40~GHz}}$\Tstrut \\\hline
        55240 & $516\pm10$ & $378\pm10$ & -- & -- & -- & -- \Tstrut\\
55698 & $407\pm9$ & $294\pm9$ & $181\pm6$ & $166\pm6$ & -- & -- \\
55848 & $408\pm5$ & $294\pm5$ & $176\pm5$ & $161\pm5$ & -- & -- \\
55873 & $397\pm8$ & $263\pm8$ & $181\pm9$ & $164\pm9$ & $128\pm12$ & $125\pm12$ \\
55892 & $373\pm10$ & $282\pm10$ & -- & -- & -- & -- \\
56075 & -- & -- & $149\pm7$ & $136\pm7$ & $94\pm9$ & $93\pm9$\\
56091 & $412\pm8$ & $286\pm8$ & $157\pm6$ & $140\pm6$ & $82\pm9$ & $80\pm9$ \\
56177 & -- & -- & $186\pm7$ & $168\pm7$ & -- & -- \\
56598 & $450\pm6$ & $329\pm6$ & -- & -- & -- & -- \\
56606 & $438\pm3$ & $308\pm3$ & -- & -- & -- & -- \\
56636 & $450\pm3$ & $312\pm3$ & -- & -- & -- & -- \\
56742 & $564\pm4$ & $410\pm4$ & $281\pm9$ & $263\pm9$ & -- & -- \\
56817 & $525\pm3$ & $410\pm3$ & $268\pm5$ & $248\pm5$ & -- & -- \\
56827 & $521\pm3$ & $406\pm3$ & $257\pm5$ & $236\pm5$ & $160\pm10$ & $155\pm10$ \\
56859 & $520\pm2$ & $375\pm2$ & -- & -- & -- & -- \\
56913 & $536\pm3$ & $395\pm3$ & $248\pm6$ & $227\pm6$ & $139\pm9$ & $134\pm9$ \\
56930 & -- & -- & $221\pm9$ & $203\pm9$ & -- & -- \\
56943 & $491\pm6$ & $363\pm6$ & $241\pm6$ & $222\pm6$ & $136\pm8$ & $131\pm8$ \\
56980 & $519\pm4$ & $379\pm4$ & $253\pm8$ & $233\pm8$ & $139\pm10$ & $133\pm10$ \\
57007 & -- & -- & -- & -- & $144\pm12$ & $138\pm12$ \\
57036 & $490\pm5$ & $372\pm5$ & -- & -- & -- & -- \\
57102 & $450\pm3$ & $330\pm3$ & $212\pm7$ & $192\pm7$ & $119\pm11$ & $114\pm11$ \\
57135 & $429\pm6$ & $292\pm6$ & $191\pm10$ & $176\pm10$ & -- & -- \\
57202 & -- & -- & $216\pm7$ & $198\pm7$ & $118\pm10$ & $115\pm10$ \\
57246 & -- & -- & -- & -- & $122\pm9$ &$119\pm9$ \\
57327 & $428\pm6$ & $293\pm6$ & -- & -- & -- & -- \\
57349 & $426\pm5$ & $306\pm5$ & $184\pm6$ & $167\pm6$ & $99\pm8$ & $95\pm8$ \\
57355 & -- & -- & $183\pm6$ & $167\pm6$ & -- & -- \\
57381 & $407\pm4$ & $295\pm4$ & -- & -- & -- & -- \\
57410 & $395\pm4$ & $268\pm4$ & -- & -- & -- & -- \\
57414 & $405\pm7$ & $302\pm7$ & -- & -- & -- & -- \\
57436 & -- & -- & -- & -- & $116\pm11$ & $112\pm11$ \\
57454 & $435\pm4$ & $316\pm4$ & $189\pm7$ & $174\pm7$ & $122\pm12$ & $121\pm12$ \\
57485 & $438\pm5$ & $327\pm5$ & $211\pm8$ & $195\pm8$ & $129\pm11$ & $126\pm11$ \\
57510 & -- & -- & $210\pm9$ & $198\pm9$ & -- & -- \\
57535 & -- & -- & $202\pm7$ & $188\pm7$ & -- & -- \\
57539 & $471\pm5$ & $328\pm5$ & -- & -- & $203\pm11$ & $205\pm11$ \\
57555 & -- & -- & $296\pm9$ & $286\pm9$ & -- & -- \\
57594 & -- & --  & -- & -- & $181\pm7$ & $176\pm7$ \\
57617 & $544\pm4$ & $409\pm4$ & $258\pm6$ & $239\pm6$ & $160\pm7$ & $157\pm7$ \\
57676 & $484\pm4$ & $405\pm4$ & $282\pm6$ & $264\pm6$ & $186\pm11$ & $184\pm11$ \\
57728 & $452\pm5$ & $333\pm5$ & -- & -- & -- & -- \\
57774 & $485\pm4$ & $362\pm4$ & -- & -- & -- & --  \\
57793 & $476\pm6$ & $340\pm6$ & -- & -- & -- & --  \\
57880 & $446\pm5$ & $313\pm5$ & $183\pm4$ & $167\pm4$ & -- & -- \\
57898 & $427\pm7$ & $306\pm7$ & -- & -- & $149\pm8$ & $146\pm8$ \\
58073 & $455\pm5$ & $358\pm5$ & -- & -- & -- & -- \\
58077 & $483\pm6$ & $374\pm6$ & -- & -- & -- & -- \\
58080 & $491\pm3$ & $378\pm3$ & -- & -- & -- & -- \\
58092 & $487\pm4$ & $398\pm4$ & -- & -- & -- & -- \\
        \hline
        \end{tabular}
\end{table*}

\begin{table*}
    \caption{Table \ref{tab:atca_data} continued.}
    \centering
    \label{tab:atca_data2}
        \begin{tabular}{lcccccc}
        \hline
        MJD & S$_{\mathrm{5.5~GHz}}$ & S$_{\mathrm{9~GHz}}$ & S$_{\mathrm{17~GHz}}$ & S$_{\mathrm{19~GHz}}$ & S$_{\mathrm{38~GHz}}$ & S$_{\mathrm{40~GHz}}$\Tstrut \\\hline
58183 & $392\pm5$ & $292\pm5$ & $205\pm8$ & $193\pm8$ & -- & -- \\
58227 & $387\pm4$ & $297\pm4$ & $194\pm7$ & $179\pm7$ & $121\pm10$ & $118\pm10$ \\
58229 & $398\pm6$ & $312\pm6$ & -- & -- & -- & -- \\
58279 & $392\pm4$ & $313\pm4$ & -- & -- & $116\pm15$ & $113\pm15$ \\
58377 & $352\pm5$ & $293\pm5$ & -- & -- & -- & -- \\
58379 & $354\pm3$ & $299\pm3$ & -- & -- & -- & -- \\
58391 & -- & -- & $254\pm5$ & $239\pm5$ & -- & -- \\
58462 & $424\pm3$ & $299\pm3$ & -- & -- & -- & -- \\
58490 & $414\pm4$ & $219\pm4$ & -- & -- & -- & -- \\
58513 & $383\pm4$ & $281\pm4$ & -- & -- & -- & -- \Tstrut\\
58562 & $433\pm5$ & $317\pm5$ & -- & -- & -- & -- \\
58590 & $408\pm3$ & $296\pm3$ & -- & -- & -- & -- \\
58601 & $425\pm7$ & $295\pm7$ & -- & -- & -- & -- \\
58715 & $442\pm3$ & $358\pm3$ & -- & -- & -- & -- \\
58727 & -- & -- & $225\pm12^{\dagger}$ & $192\pm12^{\ddagger}$ & -- & -- \\
58760 & $440\pm2$ & $347\pm2$ & $244\pm5$ & $229\pm5$  & $156\pm8$ & $152\pm8$ \\
58809 & $465\pm3$ & $352\pm3$ & $252\pm7$ & $234\pm7$ & -- & -- \\
58829 & $459\pm4$ & $327\pm4$ & -- & -- & -- & --  \\
58879 & $433\pm4$ & $332\pm4$ & -- & -- & -- & -- \\
58923 & $467\pm4$ & $362\pm4$ & -- & -- & -- & -- \\
58934 & $479\pm4$ & $358\pm4$ & -- & -- & -- & -- \\
58942 & $464\pm7$ & $414\pm7$ & -- & -- & -- & -- \\
58958 & $501\pm5$ & $359\pm5$ & $227\pm10^{\dagger}$ &  $184\pm102^{\ddagger}$ & -- & -- \\
        \hline
        \end{tabular}
\end{table*}

\end{appendix}

\end{document}